\documentclass[12pt]{article}
\usepackage[utf8]{inputenc}
\usepackage{amsfonts}
\usepackage{booktabs}
\usepackage{amssymb}
\usepackage{amsmath}
\usepackage{lscape}
\usepackage{geometry}
\usepackage{xcolor}
\usepackage{url}
\usepackage{float}
\usepackage{footmisc}
\newtheorem{theorem}{Theorem}

\newtheorem{proposition}{Proposition}

\newtheorem{assumption}[theorem]{Assumption}
\usepackage{setspace}
\usepackage{footmisc}
\usepackage{natbib}
\usepackage{apalike} 
\usepackage[final]{hyperref} 

\usepackage{enumerate}   
\usepackage{tikz}

\usepackage{accents}

\title{Kites and Quails: Monetary Policy and Communication with Strategic Financial Markets\thanks{Kites are a bird of prey and, like hawks, are in the Accipitridae family. Quail-doves are members of the Geotrygon bird genus in the pigeon and dove family. We are grateful to Simone Galperti, Aram Grygorian, James Hamilton, Michael McMahon, Denis Shishkin, Domenico Siniscalco, Joel Sobel, Pietro Spini, Johannes Wieland, and seminar participants at the CEPR Central Bank Communication Seminar Series for helpful comments and suggestions.}}
\author{Giampaolo Bonomi\thanks{Department of Economics, UC San Diego. gbonomi@ucsd.edu.} \qquad Ali Uppal\thanks{Department of Economics, UC San Diego. asuppal@ucsd.edu.}}
\date{May 2024}
\begin{document}
\maketitle
\begin{abstract}
We propose a model to study the consequences of including financial stability among the central bank's objectives when market players are strategic, and surprises compromise their stability. In this setup, central banks underreact to economic shocks, a prediction consistent with the Federal Reserve’s behavior during the 2023 banking crisis. Moreover, policymakers’ stability concerns bias investors' choices, inducing inefficiency. If the central bank has private information about its policy intentions, the equilibrium forward guidance entails an information loss, highlighting a trade-off between stabilizing markets through policy and communication. A ``kitish'' central banker, who puts less weight on stability, reduces these inefficiencies.
\end{abstract}
\newpage

\section{Introduction}
\begin{quote}
\begin{singlespace}
\textit{We might find ourselves having to act to stabilize the financial markets in order to stabilize the economy. None of us welcomes the charge that monetary policy contains a “Fed put,” but, in extremis, there may be a need for such a put, if not in the strict sense of the finance term, then at least in regard to the direction of policy. How should we communicate about such actions, or, as used to be said of the lender of last resort, should we leave such actions shrouded in constructive ambiguity?\;\footnote{\:Governor Fischer, FOMC Meeting Transcript, March 15-16, 2016 (\url{https://www.federalreserve.gov/monetarypolicy/files/FOMC20160316meeting.pdf})}}
\end{singlespace}
\end{quote}
The Global Financial Crisis of 2007-8 challenged the consensus that monetary policy should focus primarily on inflation \citep{Smets2014}. Since then, there has been a significantly revived interest in whether monetary policy should account for financial stability concerns. Indeed, \citet{Woodford2012} shows that loose monetary policy can increase financial instability and therefore central banks should include such concerns in their objective function.

In addition to such normative questions about central banks' objective functions, there are two sets of positive questions. First, do we see evidence from the revealed behavior of central banks that they might already have financial stability as part of their objective function? Second, what are the implications of this type of objective function? There is substantial evidence that central banks account for the behavior of financial markets when making decisions. We see this anecdotally from transcripts of policymakers' meetings, speeches, and other forms of communication. However, there is also more rigorous evidence of this. For example, \citet{Wilson2022} find that “the FOMC’s loss depends strongly on output growth and stock market performance and less so on their perception of current economic slack.” Moreover, there is evidence that unexpected increases in interest rates lead to a significant decline in bank stock prices as well as a decline in short-term bank profits \citep[e.g.][]{Flannery1984,Aharony1986,Alessandri2015,Busch2015,English2018,altavilla2019,Ampudia2019,Zimmermann2019,Jiang2023, Uppal2023}.

Theoretically, there appear to be at least two common approaches in formalizing ways in which central banks care about financial stability. The first relates to \emph{policy surprise}. For example, \citet{stein2018} argue that when central banks surprise the market, it can lead to costly volatility and subsequently instability for highly leveraged financial institutions. The second relates to a \emph{directional preference} of the financial market participants whereby market participants might prefer low interest rates to high interest rates. This is often related to the `Fed Put' (a term that describes a scenario in which the Federal Reserve keeps interest rates low in order to protect the stock market). \citet{Cieslak2021} highlight that the Fed Put could result in moral hazard issues.

In our paper, we seek to understand the implications of these modified central bank objective functions when financial institutions are \emph{strategic} players (i.e., they know the central bank cares about their stability). Using this game-theoretic set-up, we produce a number of contributions to the literature. First, we show that if investors (i.e., financial market participants) face costly readjustments when their investment position is not consistent with the central bank interest rate choice and the central bank internalizes part of this adjustment cost, then the central bank underreacts to shocks to the economy (see Appendix \ref{appendix:svb} for a case study using Silicon Valley Bank and Signature Bank to illuminate this underreaction result). In a sense, this is consistent with the gradualism approach highlighted by \citet{stein2018}. It is also similar to a result derived by \citet{Caballero2022}. They show that even when the Fed does not agree with the market, because market expectations affect current asset valuations, it can induce the Fed to set an interest rate that partly reflects the market's view.

Next, we focus specifically on systemic financial institutions as most central banks have explicit mandates to ensure the stability of these institutions. Such institutions are often considered ``too big to fail" and so arguably warrant additional supervision by regulators.\footnote{See \url{https://www.bis.org/bcbs/gsib/} for a list of designations.} We derive a solution to the game where these systemic institutions have payoffs that relate to policy surprise and directional preference, and central banks are concerned about market readjustment costs. In this set-up, we show that when these institutions have market power, there is a policy distortion. Indeed, the designation of the institutions as being systemic relies on them having some measure of market power. Moreover, there is considerable empirical evidence documenting the growing market power of US banks \citep[see, for instance][]{Corbae2021, Corbae2022} and increasing evidence on the role of bank market power in the transmission of monetary policy \citep{Drechsler2017}. Our market power-related distortion is akin to the moral hazard conjecture of \citet{Cieslak2021}. Put simply, given that the central bank is concerned with the stability of these institutions, they can make portfolio decisions that make it harder for the central bank to implement monetary policies misaligned with the interests of financial market participants.

We then explore two sets of policies to improve welfare. The first relates to classic papers on the appropriate degree of discretion for central banks. For example, \citet{barro1983} shows that rules for central bank behavior can be welfare-improving, but that given monetary policy is a repeated game between the policymaker and the private agents, it is possible that reputational forces might substitute for formal rules. Our findings relate more closely to the seminal paper by \citet{Rogoff1985} which finds that social welfare can be improved if the appointed central banker is more hawkish than society (i.e., a central banker who differs from the social objective function as they place a larger weight on inflation-rate stabilization relative to employment stabilization). Akin to the hawk-dove result of \citet{Rogoff1985}, we derive a similar kite-quail result. Specifically, we can show that when the central bank can be of two types: quail (place greater weight on financial institution stability) and kite (place a greater weight on the real sector, i.e., the conventional central bank objective), it is socially optimal to have a central bank that is more kitish than the society.

Finally, we contribute to the growing literature on central bank communication. While there is already a large literature on central bank communication through policy announcements, the focus is largely on communicating a decision \citep{Blinder2008}. Indeed, papers as early as \citet{stein1989} have explored the problems faced by the Fed in announcing its private information about future policies. \citet{stein1989} showed that the Fed can communicate some information about its goals through the cheap talk mechanism of \citet{CS}. However, there is much less research on the effects of central bank communication \textit{prior} to a decision being made. This is an especially important area given the prevalence of central bank speeches, press releases, and other announcements which are used to ``guide'' market expectations prior to the decision-making meeting as such communications can themselves influence the policy decision. While there is growing empirical evidence on both formal and informal communication \citep[see, for instance][]{cieslak2019, hansen2019, Bradley2020, morse2020}, as far as we are aware, only \citet{Vissing2020} explores the issue of this type of communication formally in a theoretical model. However, she does not allow for communication to have any benefit nor does she allow for the market to behave strategically, both features that seem \textit{prima facie} important in describing communication between central banks and markets. 

We find that when communication is cheap talk, a maximally \textit{kitish} central banker improves welfare more often than when the central bank can commit to full transparency. Interestingly, not only does the central bank better achieve its stability goals when a kitish central banker is appointed, but investors can also benefit, due to more transparent communication and market stability. This finding is novel and non-obvious, as the appointment of a \textit{kite} has two effects on the expected payoffs of financial market participants, pushing in opposite directions. On the one hand, it discourages attempts to influence the central banks' policy, forcing market players to behave more as ``policy takers,'' which reduces the conflict of interest with policymakers in the communication stage. As in \citet{CS}, a smaller conflict of interest allows for more information transmission, making all parties -- including market players -- better off. This benefit of a kitish central banker is \emph{in addition to} the aforementioned benefit of appointing a central banker more kitish than society, and therefore highlights a novel interaction between the central bank stance and the effectiveness of communication. On the other hand, holding the informativeness of communication fixed, financial institutions expect larger losses under a kitish central banker: the latter is less responsive to financial interests and more responsive to shocks that occur after communication, resulting in more surprises and instability. As a result, markets benefit from a central banker who puts little weight on their stability if the stability gain from transparent communication is large enough to compensate for the losses they incur due a lower policy influence.

\section{Setting}
We model the interaction between the central bank (CB) and a set of market players \(I\) using the following stylized three-period setting. At time \(t = 0\), \(CB\) privately observes a shock \(\omega_1\sim U(-\phi_1,\phi_1)\) to the economy-stabilizing policy (i.e., the interest rate consistent with stabilizing the economy), and sends a public message \(m\in M = [-\phi_1,\phi_1]\) potentially informative about \(\omega_1\). At \(t=1\), after having observed \(\omega_1\), each market player \(i\in I\) simultaneously chooses an investment position \(x_i\in X\), based on their expectation of the central bank interest rate, which is observed by the \(CB\). Finally, at \(t = 2\), \(CB\) observes a second shock to the economy stabilizing policy, \(\omega_2\sim U(-\phi_2,\phi_2)\), independent from \(\omega_1\), and chooses the policy rate \(r\in \mathbb{R}\).\footnote{The uniform distribution assumption is without loss of generality as far as \(\omega_2\) is concerned. Assuming that \(\omega_1\) is also uniformly distributed simplifies the communication analysis of section \ref{sectcom}.} The economy-stabilizing policy at the time of the decision is \(\omega\in[-\phi,\phi]\), where \(\omega = \omega_1 + \omega_2\) and \(\phi=\phi_1+\phi_2\). After the policy decision is made, the game ends and payoffs are realized.
\begin{figure}[h!]
    \centering
    \includegraphics[width=0.9\textwidth]{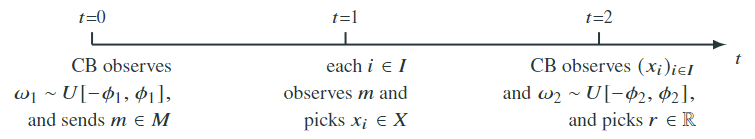}
    \caption{Timeline}
    \label{fig:time}
\end{figure}

Figure \ref{fig:time} provides a graphical representation of the timing. The interpretation of the timeline is as follows. The policy decisions of the Fed are made during meetings of the Fed Open Market Committee (FOMC), occurring once every six weeks, a period of time known as the ``FOMC cycle." We can think of our simple game as occurring within the FOMC cycle, which is distinct from much of the existing literature that considers decisions at the time of announcement. In our game, \(t= 2\) is the end of the cycle, when the policy decision is made and announced. Within the cycle, official communication occurs at fixed dates, and is meant to increase the transparency of the decision process of the FOMC, as well as provide future guidance to markets \citep[see, e.g.,][for an extensive discussion of the role of communication]{Faust2016}. With a simplifying assumption, we assume that policy-relevant communication occurs only once within the cycle, at \(t = 0\), where at the time of communication, the policymaker has some uncertainty about the economy-stabilizing policy (captured by \(\omega_2\)). This simplification is consistent with the fact that the speeches of the FOMC chair and vice-chair occur once per cycle, and that the committee observes a mandatory ``blackout period" before the occurrence of the end-of-cycle meeting. Finally, in our timeline, investors make the relevant changes to their investment positions before the policy decision \(r\) is made. This modelling choice captures the idea that before the decision is made public, at \(t=1\) market players choose their investment positions based on their future policy expectations. It allows for ``forward guidance" and highlights the economic value that investors derive from correctly predicting future monetary policies.

\subsection{Market Surprises and Underreaction}
We make two main assumptions that shape the strategic incentives of investors and policymakers in our setting. First, the payoffs of the investors are lower when the mismatch between their investment positions and the policy decision \(r\) is greater. Second, the central bank internalizes at least part of this market adjustment cost.

\begin{assumption}[Costly Market Readjustments]\label{ass1}
Let  \(X\) = \(\mathbb{R}\) and \(x_i\) denote the element of \(X\) chosen by \(i\). For \(a,b\in\mathbb{R}\), let \(S(a,b)=(a-b)^2\).\footnote{We use the quadratic loss functional form for tractability as well as consistency with the form of central bank loss functions in the literature.} We assume that
\begin{enumerate}[(i)]
    \item the payoff function of investor \(i\in I\) takes the form
    \begin{equation*}
    u_i(x_i,r)=f^i(S(x_i,r),r)    
    \end{equation*}
    for \(f^i:\mathbb{R}^2\to\mathbb{R}\) with \(f^i_1<0\), where \(f^i_1\) is the derivative of \(f^i\) with respect to its first argument;
    \item the loss function of the central bank takes the following form. For \(\alpha\in(0,1)\),
    \begin{equation*}
    L_{CB}(x,r,\omega)=\frac{1}{2}(1-\alpha)S(r,\omega)+\frac{1}{2}\alpha g(\bar{S}(x,r)),
    \end{equation*}
    for \(g:X^I\times\mathbb{R}\to\mathbb{R}\) with \(g^\prime> 0\), \(g^{\prime\prime}\ge 0\)  and \(\bar{S}(x,r)\) a weighted average over \(\{S(x_i,r)\}_{i\in I}\). 
\end{enumerate}
\end{assumption}
The assumption has a straightforward interpretation. First, (i) means that the payoff of investor \(i\) is, \textit{ceteris paribus}, decreasing in the mismatch \(S(x_i,r)\) between her investment position \(x_i\) and the policy \(r\). We can think of \(x_i\) as representing a choice from a continuum of portfolios of financial securities (e.g., stocks, bonds, and derivatives), where portfolio \(x_i = r\) is the one performing best at \(t = 2\) if the policy is \(r\). At \(t=1\), \(S\) is minimized in expectation when \(i\) holds a position \(x_i\) matching the expected future rate, and, when the investor picks a portfolio consistent with a given expected future rate, the further the realized rate is from this expectation, the worse their portfolio performs ex-post.\footnote{We simply assume that this ex-post loss is increasing in the mismatch between their portfolio and the realized policy rate in a quadratic fashion.} In this sense, \(S\) is akin to a \textit{policy surprise}. We assume that this type of portfolio performance enters the payoff function of market players, picking up an incentive for markets to correctly predict future monetary policies.

Second, assumption \ref{ass1}(ii), when paired with \(\alpha>0\), means that the central bank internalizes the cost of the overall market readjustment due to a policy surprise. The interpretation is analogous to \citet{stein2018}: unexpected policies can create large fluctuations in asset prices, undermining the financial health of highly leveraged institution.\footnote{There is a large empirical literature showing that monetary policy shocks affect asset prices (e.g., see \citet{rigobon2004} for the effects on asset prices more generally and \citet{English2018} for the effects on bank stock prices specifically).} As in \citet{stein2018} and \citet{Vissing2020}, we interpret our loss function \(L_{CB}\) as the welfare loss associated to monetary policy, so that \(\alpha>0\) means that market instability due to surprises has a negative welfare effect over and beyond the traditional channels of monetary policy.\footnote{This assumption is relaxed if we interpret \(L_{CB}\) simply as the loss of the \(CB\), and reframe the welfare implications of the next sections as payoff-implications for the \(CB\).} This assumption captures the fact that, since the Global Financial Crisis, central banks across the world have increasingly adopted explicit financial stability objectives which typically include monitoring the profitability of banks as well as the volatility of asset prices (\citet{calvo18}).\footnote{One example from the Federal Reserve comes from its May 2023 Financial Stability Report. The Fed highlighted that higher interest rates could lead to credit losses for banks exposed to commercial real estate and that as a result it would expand its monitoring of banks with such exposures. See \url{https://www.federalreserve.gov/publications/files/financial-stability-report-20230508.pdf}}\footnote{The Bank of England specifically publishes the indicators it uses for macroprudential purposes. These include the return on assets for the banking sector and the VIX index (a measure of market expectations of 30-day volatility as conveyed by S\&P 500 stock index options prices. See \url{https://www.bankofengland.co.uk/-/media/boe/files/core-indicators/countercyclical-capital-buffer.xlsx} for the full list of indicators.} While this assumption reasonably reflects central bank mandates across a growing number of jurisdictions, there is also direct empirical evidence that central banks are concerned about stock market performance (see \citet{Wilson2022} for evidence in the case of the Federal Reserve).\footnote{Although there are undoubtedly many sources of financial stability risks, we focus purely on risks to investors from unexpected changes in interest rates given the focus of this paper is on how including a financial stability objective affects the interest rate choice of the central bank.} In the next section, we explore the case when these investors are systemically important financial institutions, where the argument that the central bank is concerned for their stability is even stronger.

Not surprisingly, our first result is that when the central bank cares about market surprises, it underreacts to shocks in the economy-stabilizing policy that occur after communication. By doing so, policymakers reduce market surprises and the associated welfare cost.\footnote{This feature of the policy decision is in line with the gradualism and underreaction that characterize equilibrium policies in \citet{Vissing2020} and \citet{stein2018}, who also study monetary policy with costly policy innovations.}
This simple intuition is captured by proposition \ref{ass1}, which follows immediately from the game structure and assumption \ref{ass1}(ii).

\begin{proposition}[Underreaction]\label{underreac}
Let \(\sigma\) be any strategy profile played in a PBE of the game, and let \(r^\star(\sigma,\omega_1,\omega_2)\) be the corresponding on-path policy rate chosen by \(CB\), expressed as a function of the shocks realizations \((\omega_1,\omega_2)\). On the equilibrium path, the central bank under-reacts to the shock \(\omega_2\), that is \(0<\frac{\partial r^\star(\sigma,\omega_1,\omega_2)}{\partial \omega_2}<1\).
\end{proposition}

In the remainder of the paper, the focus of our analysis will be on a specific payoff specification that satisfies the requirements of assumption \ref{ass1} and, we believe, delivers interesting insights and policy implications.


\section{Interaction with Systemically Important Institutions}
We start by analyzing the case where the central bank cares about the financial stability of a set \(I\) of systemically important institutions. As highlighted earlier, many central banks, particularly those in advanced economies, have an explicit financial stability mandate on top of their price stability mandate. Indeed, accounting for the stability of systemic institutions is a core part of this mandate and is consistent with central banks conducting enhanced regulation and monitoring of these institutions due to the potentially systemic consequences of their instability. Moreover, central banks typically conduct stress tests of these systemic institutions to ensure that they are sufficiently stable, and in particular, that they are able to withstand adverse shocks (including unexpected changes in interest rates). 


These systemic institutions have a structural interest rate exposure related to their business model that makes their profits sensitive to unexpected changes to the direction of the policy \(r\).\footnote{We will sometimes refer to \(r\) as a policy change. This is similar to assuming that \(r=0\) at the beginning of the game.} In particular, we specify the following forms for the players' payoffs. For each \(i\in I\) and \(x_i,r\in\mathbb{R}\), 
\begin{equation}\label{payoff1}
		u_i(x_i,r)=\underbrace{-\frac{1}{2}(x_i-r)^2}_\text{policy surprise}\underbrace{-\beta r}_{\substack{\text{directional} \\ \text{preference}}}
	\end{equation}
where \(\beta>0\), implying that rate cuts positively impact investors' payoffs.\footnote{One can consider the exposure arising from maturity transformation as being determined by $\beta$ while the exposure from all other investments as being determined by the portfolio choice $x_i$. Alternatively, one can consider $\beta$ as capturing a bank's exposure to credit risk via unexpected changes in interest rates while $x_i$ captures its direct exposure to interest rate risk. In both cases, we define $\beta$ such that a positive value implies higher rates negatively impact banks.\label{footnote_beta}} We interpret \(\beta>0\) as investors having a \textit{structural} positive interest rate exposure deriving from their primary business, meaning that their profits are typically negatively impacted by rate hikes.\footnote{\citet{Begenau2020} show that the US banking sector has typically been characterized by a positive interest rate exposure over the last two decades and \citet{English2018} show that bank stock prices are negatively impacted by unexpected rate hikes.}

The main policy implications presented in this section do not depend on the sign of $\beta$.\footnote{A key difference is that a $\beta>0$ would imply that rates are declining over time while $\beta<0$ would suggest that rates are rising over time.} While the sign does not matter, we require that, on average, $\beta\ne0$.\footnote{As documented by \citet{Begenau2020}, there is heterogeneity in interest rate exposure across banks. Their results hold on average and, importantly, for large banks (market makers). In our model, we can indeed allow for heterogeneity across investors, in terms of size and sign of \(\beta\). We believe that our main results would still hold in such a richer setting, provided that \(\beta\) is on average positive (or different than 0) for systemic investors.} If $\beta$ was zero, it would imply banks have no directional preference for interest rates. While \citet{drechsler2021} show that in response to changes in interest rates, bank profits are relatively stable, when they specifically consider unexpected changes in interest rates, they find that bank stock prices actually fall. This stock price result is consistent with the findings of \citet{English2018}. In addition to a fall in bank stock prices, \citet{English2018} show that unexpected rate hikes lead to a decline in bank profits.\footnote{Much of the literature finds that unexpected interest rate hikes lead to bank stock price declines and short-term bank profit declines \citep[e.g.][]{Flannery1984,Aharony1986,Alessandri2015,Busch2015,altavilla2019,Ampudia2019,Jiang2023}} \citet{Uppal2023} finds a similar result and provides evidence that the underlying mechanism driving the overall decline in profits is due to higher credit losses from rising delinquencies.\footnote{Similarly, \citet{Zimmermann2019} finds in a repeated cross-section of 17 countries and 145 years that unexpected interest rate hikes result in greater loan losses and lower credit growth for banks, resulting in a decline in bank profits.} This latter result is consistent with considering $\beta$ as a bank's exposure to credit risk via unexpected changes to interest rates (see footnote \ref{footnote_beta}). Therefore, despite the sign of $\beta$ being inconsequential for our results, for the purposes of interpretation, we set $\beta>0$. Indeed, a positive $\beta$ is also consistent with the fact that banking crises are typically followed by reductions in the interest rate set by the central bank to, among other things, help support the banking sector.

The loss function of the central bank depends on its ability to stabilize the economy and  market instability due to a policy surprise. Specifically, for \(\omega\in[-\phi, \phi]\), \(x\in\mathbb{R}^I\) and \(r\in\mathbb{R}\),
\begin{equation}
	L_{CB}(x,r,\omega) = \underbrace{\frac{1}{2}(1-\alpha)(r-\omega)^2}_\text{conventional loss}+\underbrace{\alpha\frac{1}{2N}\sum_{i\in I}(r-x_i)^2}_\text{market readjustment}.
 \label{eq:cbloss}
\end{equation}
where \(N=|I|\). The loss function is comprised of two terms. The first term is akin to what appears in the standard objective function of New Keynesian models \citep[e.g.,][]{gali2015}. Hence, if the welfare weight of systemic investors' stability was \(\alpha = 0\), the policy \(r=\omega\) would achieve the optimal balance of inflation and economic slack. Note that this standard term already includes the change in banks' profits due to their traditional-business exposure \(\beta\), which could be considered as part of the bank lending channel of monetary policy. The second term in the central bank loss function captures the welfare cost of a financial market readjustment due to policies that surprise large investors. For simplicity, we assume that each systemically relevant institution carries equal weight in contributing to this instability (i.e., they have equal market power), so that the relevant welfare inefficiency is a simple average of the readjustment costs of the investors in \(I\).

The solution concept that we adopt for the equilibrium analysis of this simple game is Perfect Bayesian Equilibrium (PBE).\footnote{We do not need to impose additional requirements to achieve subgame perfection, since in all proper subgames \(CB\) knows all past shocks realizations.} We restrict our attention to equilibria in which the central bank choice \(r\) at \(t=2\) is not contingent on the announcement \(m\in M\) at \(t=0\). Considering equilibria in which the central bank conditions the policy choice \(r\) on previous communication would not yield additional insights, in the absence of reputational considerations. Hence, a strategy for the central bank consists of a communication rule \(\sigma_m: [-\phi_1,\phi_1]\to \Delta(M)\), mapping realizations of \(\omega_1\) to a distribution over messages in \(M\), and a policy rule \(\sigma_r:X\times[-\phi_1,\phi_1]\times[-\phi_2,\phi_2]\to\Delta(\mathbb{R})\), mapping investors' positions and realizations of \((\omega_1,\omega_2)\) to distributions over policy decisions. A strategy for market player \(i\) is an investment plan \(\sigma_{x_i}:M\to\Delta(\mathbb{R})\), mapping the central bank's messages to investment positions. We write \(\sigma_x = (\sigma_{x_i})_{i\in I}\) and denote a generic strategy profile \((\sigma_m,\sigma_{x},\sigma_r)\) by \(\sigma\in\Sigma\).

Before moving to a general solution for this game, it is useful to identify the communication strategies, investment rule and policy rule that would maximize expected welfare, as this will provide a useful benchmark for equilibrium welfare analyses.

\subsection{First Best: The Competitive Benchmark}
What would socially optimal communication, investment and policy look like? Given our assumption that \(L_{CB}\) is equivalent to the relevant measure of welfare loss in the society, it suffices to ask how the central bank would communicate, invest and set monetary policy if, similarly to a social planner, she could directly choose the investment rule \(\sigma_x\). The main result of this section is that the ex-ante optimal strategy profile is a PBE of the game where investors chose their strategies autonomously, provided that the banking sector is perfectly competitive.\footnote{In what follows, we interpret \(N\) as competitiveness of the banking sector. For instance, we can think that the central bank cares about the largest institutions that cumulatively represent some fixed share of the total market size, so that \(N\) is the number of such institutions.} Before stating this result formally and discussing it, it is useful to introduce two additional pieces of notation. First, for each \(\sigma\in\Sigma\), let \(W(\sigma)= -\mathbb{E}[L_{CB}|\sigma]\) denote the ex-ante welfare when the strategy profile is \(\sigma\). Second, we let \(\hat{W}\) denote the highest achievable welfare in the game if the central bank (or a social planner) could select any strategy profile from \(\Sigma\), that is
\begin{equation}\label{Wopt}
\hat{W} = \max\limits_{\sigma\in\Sigma}W(\sigma)
\end{equation}
and let \(\hat{\Sigma}\) be the argmax of (\ref{Wopt}).

The following proposition describes how such ``first" best strategy profiles look like, and it identifies the case where a profile in \(\hat{\Sigma}\) arises in a market equilibrium.

\begin{proposition}[Competitive Benchmark]\label{firstbest1}
\(\hat{\Sigma}\) is the set of all strategy profiles presenting each of the following three properties:
\begin{enumerate}[(i)]
\item Communication is fully informative, namely \(\sigma_m\) partitions \([-\phi_1,\phi_1]\) in singletons: different messages are sent in different states. 
\item Investment is unbiased, namely \(\sigma_{x_i}(m) = \mathbb{E}[\omega_1|m,\sigma_m]\) for each \(i\in I\) and each \(m\in M\) played with positive probability in equilibrium;
\item The central bank balances its objectives on the path of play. For \(\bar{x}\equiv\frac{1}{N}\sum_{i\in I}x_i\), and every \((\omega_1,\omega_2)\in[-\phi_1,\phi_1]\times[-\phi_2,\phi_2]\),
\begin{equation*}
\sigma_r(\omega_1,\omega_2,x) = (1-\alpha)\omega + \alpha\bar{x},
\end{equation*}
holds if \(x_i = \omega_1\) for all \(i\in I\).
\end{enumerate}
Additionally, there exist an equilibrium profile \(\sigma^{com}\) such that \((i),(ii)\) and \((iii)\) hold at the limit for \(N\to \infty\).
\end{proposition}
The proof is in the appendix. The three properties of proposition \ref{firstbest1} have intuitive interpretations: the central bank communicates all available information \(\omega_1\) about its' expected future policy, investment positions minimize expected readjustments given the expectations about the economy-stabilizing rate \(\omega\) which are formed after communication, and the final policy is chosen to optimally balance stabilization of the economy and of financial markets. Perhaps more subtly, (i), (ii) and (iii) taken together imply that in this type of game, \(\hat{W}\) does not depend on \(N\), so that looking at the mismatch between \(\hat{W}\) and the ex-ante expected welfare achievable in equilibrium for different level of competitiveness \(N\) is meaningful. 

The main implication of the proposition is that when there is perfect competition between systemic investors (i.e., \(N\to\infty\)), there is an equilibrium \(\sigma^{com}\) that achieves the maximum welfare \(\hat{W}\) that could be obtained by a benevolent planner. The intuition is as follows. First, it is easily verified that in every PBE strategy profile (including \(\sigma^{com}\)), the central bank uses the same policy rule \(\sigma^{\star}_r\), that selects a rate lying between \(\omega\) and \(\bar{x}\), optimally balancing the two stability objectives. In particular, \(\sigma^{com}_r(\omega_1,\omega_2,x)=\sigma^{\star}_r(\omega_1,\omega_2,x) = (1-\alpha)\omega + \alpha\bar{x}\), which satisfies property (iii) of proposition \ref{firstbest1}. Second, in a perfectly competitive market, investors \textit{individually} behave as policy takers. While they anticipate that policymakers will take into account the \textit{aggregate} market position \(\bar{x}\) in setting the rate, each investor is too small to influence aggregate outcomes. This implies \(\bar{x}=\mathbb{E}[\omega|m,\sigma^{com}_r]\).  To see why, note that if the average position of the market was different from the expected real economy-stabilizing policy, that is \(\bar{x}\ne\mathbb{E}[\omega|m,\sigma^{com}_r]\), then investor \(i\) would expect the central bank to follow \(\sigma^{\star}_r\) and choose an intermediate policy, so that \(\mathbb{E}[\sigma^{com}_r(\omega_1,\omega_2,x)|m,\sigma^{com}_r,x]\ne \bar{x}\). However, as policy takers, they would also find it individually rational to change their position to match perfectly this expected \(r\), minimizing the surprise loss. Hence, \(\sigma^{com}_{x_i}(m) = \mathbb{E}[\omega|m,\sigma^{com}_r]\) must hold in equilibrium, satisfying (ii). Finally, note that this ``unbiased'' investment plan is optimal also from the point of view of policymakers, who use communication to guide investors to select \(x_i = \omega_1\), the investment position that makes it easier to stabilize the economy without disrupting markets in the future. Full guidance to \(x_i = \omega_1\) is only achieved with fully transparent communication, by choosing \(\sigma^{com}_m(\omega_1) = \omega_1\) for all \(\omega_1\in[-\phi_1,\phi_1]\), or any other communication rule that satisfies (i).\footnote{Even when investors behave as policy takers, there are multiple partition equilibria with partial information transmission. Standard arguments, however, lead to the selection of the fully transparent equilibrium.}

\paragraph{Policy and Payoffs in the Competitive Benchmark} The on-path policy rate for any competitive strategy profile \(\sigma^{com}\), and in general for any \(\sigma\in\hat{\Sigma}\), is 
\begin{equation}\label{onpath1}
  r^\star(\sigma,\omega_1,\omega_2) = \omega_1 + \underbrace{(1-\alpha)\omega_2}_{\substack{\text{underreaction}}}.
\end{equation}
Transparent communication implies that the central bank fully reacts to policy target shocks occurring before communication. In contrast, and consistently with proposition \ref{underreac}, the welfare-optimal policy exhibits underreaction to state innovations that were not previously communicated, to limit market surprises. 

Simple algebra, finally, leads to the following expressions for ex-ante welfare and investors' expected payoffs, for each \(\sigma\in\hat{\Sigma}\), including the competitive limit case \(\sigma^{com}\):
\begin{gather}
W^{com}= W(\sigma) = -\frac{1}{2}\alpha(1-\alpha)\sigma^2_2\label{compwelf} \\
EU_i^{com} = \mathbb{E}[u_i|\sigma] = -\frac{1}{2}(1-\alpha)^2\sigma^2_2\label{compprof}
\end{gather}
where \(\sigma^2_2 = var(\omega_2)\). Note that both expressions are decreasing in \(\sigma_2\), proportional to the residual policy uncertainty after communication and the resulting market instability. Additionally, the first-best average welfare is the lowest when the welfare weights of real-sector instability and financial market instability are close (\(\alpha\approx\frac{1}{2}\)), while, intuitively, investors are ex-ante better-off if the central bank puts high weight on market stability.\footnote{We are not modeling here the long-term financial market losses due to an unstable economy. In this sense, we might think of our investors as ``short-term'' oriented.} 

The results of this sections seem optimistic, as the first-best average welfare can be obtained assuming competitive markets. This assumption, unfortunately, is likely to be violated in reality, where systemically important institutions -- almost by definition -- have large size and market power and there is considerable empirical evidence documenting market power in the financial sector (see e.g., \citet{Drechsler2017} and \citet{Corbae2022}). To see what happens as the banking sector becomes less competitive, we explore the general solution of the game for \(N<\infty\).

\section{Oligopolistic Competition}
When firms have market power they are typically able to influence market prices and quantities through their individual decisions. In our setting, while we did not model the economy explicitly, large systemic institutions have a similar influence over equilibrium monetary policy. To see this, note that in all PBE the central bank plays the optimal policy rule \(\sigma^{\star}_r(\omega_1,\omega_2,x) = (1-\alpha)\omega + \alpha\bar{x}\). As shown in the appendix, such a rule provides the unique minimizer of \(L_{CB}\) in any possible subgame reached at \(t=2\). Fixing this rule, for each \(i\in I\), \((\omega_1,\omega_2)\in[-\phi_1,\phi_1]\times[-\phi_2,\phi_2]\) and \(x\in X^I\) it indeed holds
\begin{equation*}
\frac{\partial\sigma^\star_r(\omega_1,\omega_2,x)}{\partial x_i} = \frac{\alpha}{N}>0
\end{equation*}
which is non-negligible when \(I\) is finite. Intuitively, when investors have market power, their \textit{individual} losses due to financial assets that perform badly under a candidate policy rate matter for aggregate market stability and therefore for the policy decision.\footnote{This argument is analogous to that in the granularity literature where idiosyncratic firm‐level shocks can explain an important part of aggregate movements given the firm size distribution (see e.g., \citet{gabaix2011}).} Hence a change in the individual position \(x_i\) affects the choice of \(r\). We refer to \(\frac{\alpha}{N}\) as the \textit{policy influence} of an individual investor.

It is worth first briefly considering whether this policy influence is plausible and importantly whether it is more likely for systemic investors (those with weight $\alpha$ in the central bank loss function) relative to non-financial corporations (those with weight $1-\alpha$ in the central bank loss function). While it is difficult to measure policy influence directly, one can certainly consider whether central banks engage more with financial institutions than non-financial institutions as this would suggest there is at least a forum for possible influence. First, central banks engage considerably more with systemic institutions given their regulatory mandate. Specifically, these institutions receive enhanced monitoring and are subject to addition regulations (e.g., the systemic risk buffer). Second, analyzing entries in the calendars of Federal Reserve governors between 2007 and 2018, \citet{morse2020} document a striking finding:  governors met with financial institutions and financial interest groups more than three times as much as with non-financial institutions and non-financial interest groups (1,573 interactions versus 507 interactions). Together, these facts provide some suggestive evidence that financial institutions have relatively more engagement with central banks and given that central banks explicitly care about their stability, financial institutions may as a result have greater policy influence.

To better understand how investors' direct policy influence changes the strategic behavior of investor \(i\), let us focus on the simple case in which the central bank has revealed \(\omega_1\) at \(t = 0\) and \(x_j = \omega_1\) for each \(j\ne i\). The expected utility of investor \(i\) at \(t=1\) given \(\sigma^{\star}_r\) and \(\bar{x}_{-i}=\omega_1\), is
\begin{align}\label{biasex}
\mathbb{E}[u_i(x_i,r)|\omega_1,\bar{x}_{-i},\sigma^{\star}_r] =  &\underbrace{- \frac{1}{2}(1-\alpha)^2\sigma_2^2 - \beta\omega_1}_{\substack{\text{expected payoff}\\
\text{from unbiased investment}}}\notag \\
&-\underbrace{\frac{1}{2}\left(1-\frac{\alpha}{N}\right)^2(x_i-\omega_1)^2}_{\substack{\text{expected readjustment due} \\
\text{to investment bias}}} -\beta\underbrace{\left[\frac{\alpha}{N}(x_i-\omega_1)\right]}_{\substack{\text{policy distortion due}\\
\text{to investment bias}}}.
\end{align} We immediately notice two differences relative to the perfectly competitive case (\(\frac{\alpha}{N}\approx 0\)). First, the expected readjustment due to investment bias is smaller, since the policy influence of \(i\) makes the central bank choose a policy rate closer to \(x_i\) than if the investor had no market power. Second, and relatedly, investment bias leads to a policy distortion that matters on top of its effect on the performance of portfolio \(x_i\), because of the positive interest rate exposure \(\beta>0\). 

This second effect is key for the next result, which shows how biased investment can be used strategically in equilibrium to distort policies in the desired direction. 

\begin{proposition}[Oligopoly Distortions]\label{olidist}
The oligopolistic equilibrium investment plan \(\sigma^{oli}_{x}\) entails investment bias. More generally, fix any communication rule \(\sigma_m\) and let \(\sigma^{oli}_{x}\) be such that each investor best responds to \((\sigma_m,\sigma^{\star}_r)\) and to all other investors strategies. Then,
\begin{enumerate}[(i)]
\item Investment decisions are consistent with a lower interest rate relative to the one expected to stabilize the economy. The bias magnitude is strictly increasing in \(\alpha\) and \(\beta\) and strictly decreasing in \(N\). For each \(i\in I\) and \(m\in M\)
\begin{equation*}
\sigma^{oli}_{x_i}(m) - \mathbb{E}[\omega_1|m,\sigma_m] = -\frac{\alpha\beta}{(1-\alpha)(N-\alpha)}.
\end{equation*}
\item If the central bank fully reveals \(\omega_1\), that is, \(\sigma_m=\sigma^{com}_m\), than the on-path policy decision exhibits a downward distortion. For each \(\omega_1\in[-\phi_1,\phi_1]\) and \(\omega_2\in[-\phi_2,\phi_2]\),
\begin{equation*}
r^\star(\sigma^{oli,tr},\omega_1,\omega_1) = r^\star(\sigma^{com},\omega_1,\omega_2) - \frac{\alpha^2\beta}{(1-\alpha)(N-\alpha)}
\end{equation*}
where \(\sigma^{oli,tr}=(\sigma_m^{com}, \sigma^{oli}_x,\sigma_r^\star)\), with \(\sigma_m^{com}\) fully informative about \(\omega_1\).
\end{enumerate}
\end{proposition}

To see why investment bias arises in equilibrium, consider again the case where the central bank always fully reveals \(\omega_1\) and in which all investors other than \(i\)  have chosen unbiased positions \(x_j = \omega_1\). The first order condition in equation \ref{biasex} is 
\begin{equation*}
\underbrace{\left(1-\frac{\alpha}{N}\right)^2(\omega_1-x_i)}_{\substack{\text{marginal cost from} \\ \text{downward bias}}} = \underbrace{\frac{\alpha\beta}{N}}_{\substack{\text{marginal benefit from} \\ \text{downward bias}}}
\end{equation*}
which yields the bias of proposition \ref{olidist} if \(N = 1\).\footnote{Recall that for the sake of developing intuition we assumed \(x_{-i}=\omega_1\). This will not be true in equilibrium, so that the expression of the equilibrium bias for a generic \(N\), reported in proposition \ref{olidist}, is different from the one we obtain from this special case.} The marginal benefit from a downward investment bias is positive for \(I\) finite, and therefore \(x_i = \omega_1\) is not optimal for an individual investor, even when all other institutions are choosing unbiased investment. And this is more true the higher the individual policy influence \(\frac{\alpha}{N}\) and the higher the benefit from a rate cut \(\beta\). It is easy to see that in equilibrium, each systemic institution will choose portfolio positions consistent with smaller rate hikes than those really expected, potentially increasing their own losses from a restrictive change in monetary policy. This result, apparently surprising, is consistent with the evidence by \citet{Begenau2020}, who find that the 4 largest US banks typically hold net derivative positions reinforcing (instead of hedging) their interest rate exposure.

Why would investors take risky positions that create even more losses in case of adverse rate shocks? The key is that, by behaving as if the rates were to remain lower than optimal for the economy, the systemic institutions in our game make it harder for the central bank to make large and undesired policy changes. This is the result shown in the second part of proposition \ref{olidist} if we assume commitment to fully informative communication: the on-path rate decision is systematically lower than the one that maximizes welfare. As such, our results provide one possible microfoundation for the empirical evidence on interest rate exposure.

Moreover, our results are similar in spirit to some of the findings in the literature around risk-taking of `too-big-to-fail' institutions (which we refer to as systemic institutions in this paper). For example, \citet{afonso2014} show empirically that banks deemed too-big-to-fail engage in greater risk-taking because they believe they will be rescued if they fail. However, taking greater risk, all else equal, increases the likelihood of their failure and so pushes the regulatory authority to provide more protective measures (e.g., asset guarantee programs) than it would have otherwise. Similarly, in our paper, systemic institutions increase their exposure to losses from higher rates which pushes the central bank to be more accomodative by cutting rates or raising rates less than it would have otherwise. A recent example that highlights the pressure the Federal Reserve faces in relation to raising interest rates was during the banking panic of 2023. During this time, the former chair of the Federal Deposit Insurance Corporation (one of the institutions with a financial stability mandate in the US) publicly called on the Federal Reserve to stop raising interest rates.\footnote{See \url{https://www.cnn.com/2023/03/13/investing/sheila-bair-svb-fed-rates/index.html}} Indeed, during this banking panic, both the Fed's behavior and market expectations were consistent with the underreaction prediction of our model (see Appendix \ref{appendix:svb} for details). As we will see in the next section, a similar result holds, \textit{ex-ante}, in the strategic communication equilibrium.

A consequence of these distortions is that when the central bank commits to full information transmission at \(t=0\) and players best respond, the society incurs an ex-ante welfare loss relative to perfect competition. In particular, it holds 
\begin{equation*}
    W(\sigma^{oli,tr}) - W(\sigma^{com}) = -\frac{\alpha^3\beta^2}{2(N-\alpha)^2(1-\alpha)}<0.
\end{equation*}
Investors benefit from their policy influence, achieving a higher ex-ante payoff than in the perfectly competitive case. Indeed, it can be shown that, for each \(i\in I\),
\begin{equation*}
EU^{oli,tr}_{i} - EU^{com}_i = \frac{(2N-1-\alpha)\alpha^2\beta^2}{2(N-\alpha)^2(1-\alpha)}>0. 
\end{equation*}
As one would expect, both these gaps close if the degree of competition \(N\) increases, if the welfare weight \(\alpha\) of market stability decreases, or if the structural exposure of systemic investors \(\beta\) increases.

We have shown that when the market is not perfectly competitive, investment is generally biased, and, in the case when the central bank communicates all available information about the economy-stabilizing policy, there is, on average, a welfare loss for the society. Can more communication flexibility close this gap relative to perfect competition? Or will it instead lead to further welfare losses? We address this question in the next section.

\subsection{Communication}\label{sectcom}
We now relax any assumption about commitment to transparent communication and look at how equilibrium strategic communication will look like in our simple model, and what its welfare implications will be. An unfortunate implication of proposition \ref{firstbest1} and proposition \ref{olidist} is that communication cannot bring us back to the ex-ante optimum \(W(\sigma^{com})\). In fact, in any oligopolistic PBE, investment is biased downwards relative to \(\mathbb{E}[\omega_1|m,\sigma^{oli}_m]\), the market expectations of the future economy-stabilizing rate given any equilibrium communication rule \(\sigma^{oli}_m\) and the announcement \(m\) made at \(t = 0\). This means that the unbiasedness requirement of proposition \ref{firstbest1} is violated so that \(\sigma^{oli}\notin\hat{\Sigma}\). Even more importantly, given the cheap talk nature of comunication without commitment in this environment, flexibility ends up creating an additional welfare loss relative to commitment to transparent communication. 

It is easy to show that equilibrium communication has the exact same characteristics as the communication strategy of the seminal cheap talk game studied by \citet{CS}, where the receiver sender bias \(b = \frac{\alpha\beta}{2\phi_1(N-\alpha)(1-\alpha)}\). 

\begin{proposition}[Cheap Talk]\label{communication}
Any communication rule played in an oligopolistic PBE partitions \([-\phi_1,\phi_1]\) in a finite number of intervals, so that equilibrium communication is never fully informative. In addition: 
\begin{enumerate}[(i)] 
\item A PBE strategy profile \(\sigma^{oli}_{P}=(\sigma^{oli}_{m,P},\sigma^{oli}_x,\sigma^{\star}_r)\) corresponding to \(P\) partition elements, produces the following ex-ante welfare:
\begin{equation*}
W(\sigma^{oli}_P) = W(\sigma^{oli,tr}) - \frac{1}{2}\alpha(1-\alpha)\hat{\sigma}^2_{1,P}
\end{equation*}
where \(\hat{\sigma}^2_{1,P}\) is the residual variance of \(\omega_1\) induced by the equilibrium communication strategy. 
\item The residual variance \(\hat{\sigma}^2_{1,P}\) is decreasing in the absolute value of the investment bias, and in the number of partition elements \(P\). The maximum number of partition elements \(\bar{P}\) is the smallest integer greater than or equal to
\begin{equation*}
\frac{1}{2}\left(\sqrt{1+\frac{4\phi_1(N-\alpha)(1-\alpha)}{\alpha\beta}} -1\right)
\end{equation*}
which is non-decreasing in the absolute value of the investment bias.
\end{enumerate} 
\end{proposition}
As shown in the appendix, proposition \ref{communication} follows immediately from \citet{CS} and from the game structure once we impose individual rationality of investment and policy decisions. 

The main implication of the result is that fully informative communication about policy intentions would be optimal for the central bank in this context, since welfare is decreasing in any the residual uncertainty about \(\omega_1\), as described in point (i). However, unfortunately, a fully informative equilibrium is not achievable. To see this, imagine that the \(CB\) was playing \(\sigma_m(\omega_1) = \omega_1\) in a PBE, which is fully informative. Investors' would best respond with an investment plan \(\sigma^{oli}_x(m) = m - \frac{\alpha\beta}{(N-\alpha)(1-\alpha)}\), to bias the future rate downwards. But the central bank would then be tempted to systematically announce higher rate hikes at \(t=0\), deviating to \(\sigma_m(\omega_1)=\omega_1+\frac{\alpha\beta}{(N-\alpha)(1-\alpha)}\). This credibility loss arises from the fact that the interests of the central bank and investors are not fully aligned and, when investors have policy influence, they do not behave as policy takers. 

The consequence of the loss in information transmission relative to commitment to informative communication is that, on average, the central bank will provide less guidance, markets will be less ready for changes in policies, and policies will create more instability. This negative effect is minimized in the most informative equilibrium, the one with \(P=\bar{P}\), since more partitions imply more informative communication. Not surprisingly, point (ii) of proposition \ref{communication} shows that the informativeness of communication in this society-preferred equilibrium increases the closer the investment distortion is to zero. Hence, the maximum amount of guidance achieved in equilibrium by the central bank increases in the competitiveness of the banking sector and decreases in the welfare weight of market surprises and in exposures of investors.

Note that investors too are ex-ante worse-off when the central bank does not communicate clearly. When the equilibrium strategy profile is \(\sigma^{oli}_{P}\), the systemic investors' loss relative to commitment to full information is 
\begin{equation}
EU_i^{oli,tr} - EU_i^{oli,P}= \frac{1}{2}(1-\alpha)^2\hat{\sigma}^2_{1,P}>0
\end{equation}
in fact, being less able to predict future policies, they will be find it harder to balance the need to follow the central bank and the one to influence it.

Finally, notice that, for any equilibrium with \(P\le\bar{P}\) the on-path policy decision distortion will be negative on average, 
\begin{equation*}
\mathbb{E}[r^\star(\sigma_P^{oli},\omega_1,\omega_2)-r^\star(\sigma^{com},\omega_1,\omega_2)] = -\frac{\alpha^2\beta}{(N-\alpha)(1-\alpha)}
\end{equation*}
but positive for some realizations of \(\omega_1\) (e.g., those very close to the upper bound of the respective partition element). This is due to the partial information transmission of the cheap talk equilibrium.

One of the key insights of this section is that the central bank faces a trade-off between avoiding surprises \textit{ex-post} and communicating effectively: on the one hand, market stability concerns (\(\alpha>0)\) push policymakers to systematically underreact to state innovations \(\omega_2\) that could surprise markets. Such ex-post policy adjustments can be desirable for the reasons discussed in our first-best analysis: market surprises generate welfare-detrimental instability. On the other hand, the greater the stability concerns \(\alpha\), the larger the investment bias and the communication loss: the central bank's ex-post effort to avoid policy surprises reduces the effectiveness of forward guidance, a force that increases instability.

\subsection{Towards a Kitish Central Bank}\label{sectkites}
We have shown that systemically important investors can use market power to influence the policies of the central bank, and that this creates an inefficiency and a welfare loss, especially when the central bank has private information that cannot be communicated credibly. What remedies could we use to fix these issues? 

In the beginning of our discussion of the oligopoly, we have introduced the concept of policy influence, which we measured by \(\frac{\alpha}{N}\) the direct effect of an individual investor behavior on the central bank choices. A positive policy influence is what makes the oligopoly different from perfect competition, and, indeed, what drives the inefficiency due to investment and policy distortions. An obvious way to do so would be reducing the scale of systemically important institutions. This would likely be a long and difficult process, that might create inefficiencies due to economies of scale, not modeled here, and indeed as mentioned earlier, concentration appears to be increasing over time. The second possibility would be to reduce \(\alpha\). While we cannot reduce the \textit{welfare weight} of a market disruption, we can consider what would happen if we were to appoint a (representative) central banker with an objective function different from societal welfare. In particular, in a spirit similar to \citet{Rogoff1985}, we conduct the following thought experiment: imagine that we could appoint a central banker who has an objective with the same functional form as in (\ref{eq:cbloss}), but with a relative weight \(\tilde{\alpha}\in[0,1]\) of market stability potentially different from \(\alpha\). What would the optimal central banker look like?

Formally, let us denote by \(\sigma^{oli,tr}(\tilde{\alpha})\) and \(\sigma^{oli}_{\bar{P}}(\tilde{\alpha})\), respectively, the oligopolistic strategy profile under transparency and the cheap talk oligopolistic equilibrium when the weight in the loss function of the central bank is \(\tilde{\alpha}\). We then ask what is the level of \(\tilde{\alpha}\) that maximizes \(W(\sigma^{oli,tr}(\tilde{\alpha}))\) or \(W(\sigma^{oli}_{\bar{P}}(\tilde{\alpha}))\), where the function \(W\), as in the previous section, maps strategy profiles into ex-ante welfare levels (computed based on the welfare weight \(\alpha\)).

To better illustrate the answer, let us denote a central banker with \(\tilde{\alpha}<\alpha\) as \textit{kitish} and a central banker with \(\tilde{\alpha}>\alpha\) as \textit{quailish}. Finally, we denote a central banker with preferences equal to the society (\(\tilde{\alpha}=\alpha\)), as unbiased. A kitish central banker focuses on the real economy, paying relatively litle attention to market positions, while a quailish central banker is more accommodating towards financial markets. Our main result is that it would be optimal for the society to appoint a kitish central banker. 

\begin{proposition}[Kites]\label{kites}
The optimal central banker is kitish. In particular, the following holds:
\begin{enumerate}[(i)]
\item There exist thresholds \(\sigma_2^{oli}, \sigma_2^{oli,tr}>0\) on the non-communicable uncertainty such that: under transparency, maximally kitish central bankers (\(\tilde{\alpha} = 0\)) improve welfare over unbiased ones if and only if \(\sigma_2<\sigma_2^{oli,tr}\); under cheap talk, they improve over unbiased ones if and only if \(\sigma_2<\sigma_2^{oli}\). Moreover, it has \(\sigma_2^{oli} > \sigma_2^{oli,tr}\).
\item The optimal central bankers are such that \(\tilde{\alpha}^{oli}, \tilde{\alpha}^{oli,tr} \in (0,\alpha)\), where \(\tilde{\alpha}^{oli,tr}\) is the central banker under transparency and \(\tilde{\alpha}^{oli}\) is the central banker under cheap talk. The welfare loss under transparency and \(\tilde{\alpha}^{oli,tr}\) is smaller than the loss under cheap talk and \(\tilde{\alpha}^{oli}\). Additionally, \(\tilde{\alpha}^{oli,tr}\) strictly positive and increasing in \(\alpha\), \(N\), and \(\phi_2\).
\item  When the central bank commits to transparency, the market ex-ante payoff is always strictly larger when the central banker is unbiased than when she is kitish. However, if communication is cheap talk, then the market ex-ante payoff can be larger under a kitish central banker than under an unbiased one.  
\end{enumerate}
\end{proposition}

The intuition for why society might prefer a kitish central banker is simple. As discussed above, a kitish central banker reduces systemically relevant investors' influence over policies, and is more welfare-enhancing the larger the market power of investors. If the choice is between an unbiased central banker and one that assigns no weight on market stability, part (i) tells us that the latter will be desirable when state innovations occurring after communication are of limited magnitude (\(\sigma_2^2\) low), especially when a kite can also improve information transmission (i.e., under cheap talk). If \(\tilde{\alpha}\) can be fine-tuned, the optimal central banker is always kitish, but not maximally so: the reason why the optimal weight is non-zero is that market surprises are costly for society (\(\alpha>0\)), so that some degree of market stabilization is always optimal. 

What is perhaps more surprising is the difference highlighted by part (iii) of the proposition, that is, under cheap talk a kite can even improve the performance of financial institutions. The key intuition is that when misalignment of incentives compromises information transmission (in the cheap talk equilibrium), a more kitish central banker improves communication: a kitish central banker reduces the marginal net benefits from investment downward bias, making it optimal for the investors to choose a portforlio position \(x_i\) more aligned with their true expectations about \(\omega\). Consistently with proposition \ref{communication}.ii, this bias reduction means that more information is communicated in equilibrium, and the resulting reduction in policy uncertainty has a stabilizing effect on financial markets, making \textit{both} the society and the financial institutions better off. If this gain in information transmission is large enough, market players are willing to give up some of their policy influence to achieve it (as is the case when a kitish central banker is appointed). The following example provides a stark illustration of this phenomenon.

\paragraph{Babbling Monopoly}
Let \(N=1\), and \(\phi_1=\frac{1}{2}\). It is easy to verify that in this monopolistic setting \(\frac{\alpha\beta}{(1-\alpha)^2}>\frac{1}{4}\) implies that \(\bar{P}=1\), so that \(\hat{\sigma}^2_{1,\bar{P}}=\frac{1}{12}\) in the cheap talk equilibrium. Under an unbiased central banker, the monopolistic investor's ex-ante value of the game is \(EU^\alpha = -\frac{1}{2}(1-\alpha)^2\left[\frac{1}{12}+\sigma_2^2+\frac{(\alpha\beta)^2}{(1-\alpha)^4}\right]+\left(\frac{\alpha\beta}{1-\alpha}\right)^2\), which simplifies to \(-\frac{1}{2}(1-\alpha)^2\left[\frac{1}{12}+\sigma_2^2-\frac{(\alpha\beta)^2}{(1-\alpha)^4}\right]\). With a maximally kitish central banker, i.e., \(\tilde{\alpha}=0\), the investors' ex-ante value of the game is \(EU^0 = -\frac{1}{2}\sigma_2^2\) in the fully informative equilibrium. If \(\frac{1}{12}-\frac{(\alpha\beta)^2}{(1-\alpha)^4}>\frac{1-(1-\alpha)^2}{(1-\alpha)^2}\sigma_2^2\) then \(EU^0>EU^\alpha\), and our monopolistic investor would give up all its policy influence to improve communication. Note that the condition is satisfied for \(\frac{\alpha\beta}{(1-\alpha)^2}\in(\frac{1}{4},\frac{1}{\sqrt{12}})\) and \(\sigma_2^2\) small enough.
\subsection{Game Repetition}
In this section we return to the assumption that \(\tilde{\alpha} = \alpha\), and we consider an infinite-horizon repetition of the game outlined in the previous section. Let \(\tau = 0,1,...\) index the stage game repetition, and let \(\omega^{\tau}, m^{\tau}, x^{\tau}, r^{\tau}\) denote the respective quantities in the stage game \(\tau\). 

We address the following question: can an infinitely repeated interaction allow the central banker to better discipline markets, communicate more transparently and achieve the first best? We decide to focus on equilibria such that the investment decision \(x^{\tau}_i\) is contingent only on past messages \((m^0,...,m^\tau)\) and investment decisions \((x^0_i,...,x^{\tau-1}_i)_{i\in I}\), and is therefore independent from central bank past policies, or shock realizations. \footnote{The set of equilibria of the repeated game depends on the observability of the shocks \(\omega_1\) and \(\omega_2\) at the end of the stage game. In particular, when the central bank equilibrium policies are contingent on shock realization, any market strategy treating histories where the central bank has deviated differently from histories where the central bank has played according to the equilibrium requires market players to draw inferences about actual shock realizations to detect such deviations. The relation between \(m^{\tau}\) and \(r^{\tau}\) can sometimes convey information about a central bank deviation, we decide to abstract from these situations to simplify the analysis and avoid imposing additional assumptions.}

We will see in the next propositions that, even within the restricted class of equilibria, repeated interaction can indeed achieve the first best on the path of play with a very simple type of PBE strategy profile, provided that forward guidance is valued enough by markets (i.e., \(\phi_1\) is high). At the same time, repeated interaction can facilitate collusion between large investors, sometimes with perverse consequences on welfare.

\begin{proposition}[CB Disciplines Markets]\label{dynfirstbest}
Consider the infinitely repeated game with discount factor \(\delta\in(0,1)\). If \(\phi_1 > \frac{\alpha\beta}{(N-\alpha)(1-\alpha)}\sqrt{3\frac{2N-\alpha-1}{1-\alpha}}\) then there exist a threshold \(\delta^\star\in (0,1)\) such that if \(\delta\ge\delta^\star\) the ex-ante stage welfare is at its first-best level on the equilibrium path of some PBEs. Such PBE strategy profiles are qualitatively equivalent to the following simple profile. 
\begin{enumerate}[(i)]
\item At \(\tau = 0\), \(CB\) selects \(m^{\tau} = \omega^\tau_1\). At \(\tau>0\), \(CB\) selects \(m^\tau = \omega^\tau_1\) if \(x^s_i = m^s\) for each \(s<\tau\) and each \(i\in I\), while \(CB\) draws \(m^\tau\sim U(-\phi_1,\phi_1)\) after any other history.

\item At \(\tau = 0\), each \(i\in I\) selects \(x^\tau_i = m^\tau\). At \(\tau > 0 \) each \(i\in I\) selects \(x^\tau_i = m^\tau\) if \(x^s_j = m^s\) for each \(s<\tau\) and each \(j\in I\), while each \(i\in I\) selects \(x^\tau_i = -\frac{\alpha\beta}{(N-\alpha)(1-\alpha)}\) after any other history.

\item At each \(\tau = 0,1,...\), \(CB\) sets rate \(r^\tau = (1-\alpha)\omega^{\tau} + \alpha\bar{x}^\tau\).
\end{enumerate}
\end{proposition}

The proposition has a very simple interpretation. If the variance of the shock \(\omega_1\) is large enough (\(\phi_1\) high), the value of forward guidance is high for market players. The central bank can leverage its informational advantage by conditioning fully informative guidance on market discipline -- threatening to revert to meaningless communication if the directives are not followed in previous periods. The threat is credible, because when investors expect communication to be uninformative and stop listening, no deviation from the central bank can restore confidence in its announcements. In this equilibrium, policymakers play an active role in disciplining large banks. After investors' deviations, the central bank does not simply revert to the most efficient stage PBE communication rule presented section \ref{sectcom}: market discipline follows from a threat to revert to maximally inefficient communication. This threat allows for market discipline, even in cases where investors would be better off under repetition of the stage game equilibrium \(\sigma^{oli}_{\bar{P}}\). 

It seems natural, at this point, to ask whether the first best can also be sustained by spontaneous coordination of market players, with policymakers playing a less active role. To address this question, we restrict the attention to equilibria that satisfy two additional requirements. First, we impose that the central bank communicates as transparently as possible given the expected equilibrium investment bias \(\mathbb{E}[\bar{x}^\tau - \omega_1^\tau|x^0,...,x^{\tau-1}, m^0,..., m^{\tau-1}]\) at the history reached. The focus on most informative equilibrium communication in every contingency can be seen as the dynamic equivalent of what we did in section \ref{sectcom} in a static setting. Second, we focus on equilibria where the ex-ante expected stage payoff of each investor \(i\in I\) is weakly greater than the expected payoff of the static game equilibrium \(\sigma^{oli}_{\bar{P}}\). We call these equilibria \textit{collusive}, as they represent a (weak) Pareto improvement for large investors relative to the static setting (where there is no coordination), and the central bank plays a passive role in disciplining markets. 

The next proposition suggests that under certain conditions there exist equilibria where the oligopolists coordinate on the competitive equilibrium, but repeated interaction can also lead to detrimental effects on stability of the financial markets and the real economy. 

For each \(n = 1, ...., N\), and fixed parameters \(\alpha,\beta,\phi_1,\phi_2\), let \(\hat{\sigma}^2_{1,\bar{P}}(n)\) denote the residual variance of \(\omega_1\) after communication in the most informative stage-game equilibrium when the market consists of \(n\) large investors.

\begin{proposition}[Market Collusion]\label{coord}
Consider the infinitely repeated game with discount factor \(\delta\in(0,1)\). The following holds:
\begin{enumerate}[(i)]
    \item If \(\hat{\sigma}^2_{1,\bar{P}}(N)>\left[\frac{\alpha\beta}{(N-\alpha)(1-\alpha)}\right]^2\frac{2N-\alpha-1}{1-\alpha}\), then there exist \(\delta^\star_1\in (0,1)\) such that, if \(\delta\ge\delta^\star_1\), the first-best ex-ante welfare is the ex-ante stage welfare of some collusive PBE of the infinitely repeated game.
    \item If \(\hat{\sigma}^2_{1,\bar{P}}(1)-\hat{\sigma}^2_{1,\bar{P}}(N)<\left[\frac{\alpha\beta}{(1-\alpha)^2}\right]^2-\left[\frac{\alpha\beta}{(N-\alpha)(1-\alpha)}\right]^2\frac{2N-\alpha-1}{1-\alpha}\), then there exist \(\delta^\star_2\in (0,1)\) such that, if \(\delta\ge\delta^\star_2\), the ex-ante welfare of the monopolistic static game is the ex-ante stage welfare of some collusive PBE of the infinitely repeated game.
\end{enumerate}
Investors are ex-ante better-off in the monopolistic equilibrium (ii) relative to the first best (i) if \(\hat{\sigma}^2_{1,\bar{P}}(1)<\left[\frac{\alpha\beta}{(1-\alpha)^2}\right]^2\), while they are better off at the first best if \(\hat{\sigma}^2_{1,\bar{P}}(1)>\left[\frac{\alpha\beta}{(1-\alpha)^2}\right]^2\).
\end{proposition}

The first part of the proposition suggests that if the policy uncertainty that can be resolved by forward guidance is sufficiently large -- hence payoff-relevant for investors -- then market players will benefit from self-discipline: investors refrain from using their market power to bias future policies, which allows the central bank to communicate all its information on future policy changes, achieving the first best. 

The second part of the proposition serves as a caveat: unsurprisingly, collusion between investors might well push in the opposite direction, facilitating a stronger exercise of market power by strategic market players, allowing them to act collectively as a single large player. 

It is possible that large institutions might try to coordinate on investment strategy that maximize their aggregate expected payoffs. The type of collusion that is optimal for large investors need not be efficient from the central bank perspective, nor need it be as inefficient as the monopolistic one. However, we believe that the two equilibria highlighted in proposition \ref{coord} are important focal points from the point of view of the analyst and, potentially, the policymaker. The end of proposition \ref{coord} provides a sufficient condition for the monopolistic collusive equilibrium to benefit large investors more than the first best. If the gains from the exercise of monopolistic market power are large enough, markets should not be expected to self-coordinate on the first best, even when the first best is among the feasible collusive PBE.

All in all, the (partial) analysis of the repeated game suggests that, dynamic incentives could lead to both higher or lower central bank losses. To maximize the chances of achieving efficient outcomes, policymakers might have to play an active role, threatening to withhold future guidance if large market players refuse to cooperate.\footnote{Extending the analysis to broader set of equilibria, including those where market strategies are contingent on previous monetary policy, would likely lead to further interesting implications. Large investors -- and not only policymakers -- could strategically exploit communication to further bias policy or increase the informativeness of announcements. For instance they could threaten the central bank to ignore future forward guidance if previous announcement are not precise enough or previous rate changes are not accommodating enough.}

\section{Discussion}
We have shown that when the central bank cares about market stability and systemic investors, like large banks, are asymmetrically affected by rate hikes and rate cuts, the latter can use their market power to make adverse rate changes less likely. Specifically, they can choose market positions that would create a costly readjustment if the central bank were to choose a high economy-stabilizing rate (i.e., a high policy rate). Intuitively, this policy distortion depends on investors' policy influence \(\frac{\alpha}{N}\) which is decreasing in the degree of competitiveness of the market \(N\), and increasing in the welfare weight of market instability \(\alpha\). The resulting welfare loss is even larger if the central bank cannot guide markets credibly, which is reasonable when the misalignment between markets and policymakers makes the former mistrustful towards announcements. Increasing competition in financial markets (i.e., transitioning towards a larger number of smaller and ``less systemic" institutions) and appointing a central banker who puts little weight on market stability might increase both information transmission and welfare in equilibrium. Repeated interaction is not guaranteed to resolve the conflict of interest: the central bank can try to discipline markets under the threat of withholding future guidance if the strategic investors exercise their policy influence. But, when forward guidance is not valuable enough, market discipline might not be feasible, and large institutions could instead increase their effective market power through oligopolistic collusion. 

The simple analysis yields a number of predictions. It suggests that a central bank that cares about large investors is expected to tailor policies to the interest rate exposures of large investors, with this tailoring increasing in the concentration of the financial market. As a consequence, net interest rate exposures of the largest systemic investors should systematically predict the variation in interest rate choices not explained by economic fundamentals. Moreover, the model would also predict that the growing market power over time should result in greater underreaction by the central bank and, all else equal, a declining trend in the interest rate. Indeed, while not a causal claim, we do see that in the data rising bank market power since the 1980s has coincided with a steadily declining interest rate. While the effectiveness and informativeness of forward guidance are in general expected to be limited when systemic investors have large market power, market discipline and transparent communication should be more common when the uncertainty regarding economic fundamentals is sufficiently high.
\newpage


\bibliographystyle{apalike}
\bibliography{refs}
\newpage
\appendix
\section{Appendix: Proofs}
\paragraph{Proof of proposition \ref{underreac}}
Assume that a PBE of the game exists and let \(\sigma = (\sigma_m,\sigma_x,\sigma_r)\) be the corresponding strategy profile, consisting of a communication rule \(\sigma_m:[-\phi_1,\phi_1]\to \Delta(M)\), a rate rule \([-\phi_1,\phi_1]\times[-\phi_1,\phi_1]\times X^I\to\Delta(\mathbb{R})\) for \(CB\), and an investment plan \(x_i: M\to \Delta(X)\) for each \(i\in I\).\footnote{We only look at rate rules which are not contingent on messages \(m\in M\), as doing otherwise is not meaningful in this static setting.} Fix any \((\omega_1,\omega_2,m,x)\in [-\phi_1,\phi_1]\times[-\phi_2,\phi_2]\times M\times X^I\), and consider the problem of the \(CB\) at \(t=2\) in the subgame identified by \((\omega_1,\omega_2,m,x)\). In any PBE, the \(CB\) must pick \(\hat{r}\) solving 

\begin{equation*}
\min_{r\in \mathbb{R}} L_{CB}(x,r,\omega)
\end{equation*}
for \(\omega = \omega_1+\omega_2\). Taking the derivative of the objective function with respect to \(r\) one obtains 

\begin{equation*}
\frac{\partial L_{CB}}{\partial r} = (1-\alpha)(r-\omega) + \alpha g^{\prime}(\bar{S}(x,r))(r-\bar{x})
\end{equation*}
for \(\bar{x}\) a weighed average of \(x_1,...,x_I\). Given assumption \ref{ass1}, the first order condition is necessary and sufficient for optimality, and gives the unique PBE rate choice \(\sigma^{\star}_r(\omega_1,\omega_2, x)\) made by \(CB\) in any subgame \((\omega_1,\omega_2, m, x)\) reached at \(t=2\) as a function of the corresponding \((\omega_1,\omega_2, x)\). From the first order condition, \(\sigma^{\star}_r(\omega_1,\omega_2, x)\) must satisfy

\begin{equation}\label{sstar}
 \sigma^{\star}_r(\omega_1,\omega_2, x) = \frac{1-\alpha}{(1-\alpha)+\alpha g^\prime(\bar{S}(x,r))}\omega + \frac{\alpha g^\prime(\bar{S}(x,r))}{(1-\alpha)+\alpha g^\prime(\bar{S}(x,r))}\bar{x}.
\end{equation}
Since this must hold in any PBE, it has \(\sigma_r = \sigma^{\star}_r\).
Note that communication and investment decisions are not functions of \(\omega_2\), since they are taken before \(\omega_2\) is realized. Hence for each \((\omega_1,\omega_2)\in [-\phi_1,\phi_1]\times[-\phi_2,\phi_2]\) and \(m\in \mathrm{supp}(\sigma_m(\omega_1))\), and \(x_i\in\mathrm{supp} 
 (\sigma_{x_i}(m))\),
\begin{equation*}
\frac{\partial \sigma^{\star}_r(\omega_1,\omega_2,x)}{\partial \omega_2} = - \frac{\frac{\partial^2 L_{CB}}{\partial \omega_2\partial r}}{\frac{\partial^2 L_{CB}}{\partial r^2}} = \frac{1-\alpha}{(1-\alpha)+\alpha \left[2g^{\prime\prime}(\bar{S}(x,r))(r-\bar{x})^2+g^{\prime}(\bar{S}(x,r))\right]},
\end{equation*}
so that \(0<\frac{\partial \sigma^{\star}_r}{\partial \omega_2}<1\) follows from \(\alpha\in(0,1)\), \(g^\prime>0\) and \(g^{\prime\prime}\ge 0\). Hence \(0<\frac{\partial r^{\star}(\sigma,\omega_1,\omega^2)}{\partial \omega_2}<1\) for each \((\omega_1,\omega_2)\in [-\phi_1,\phi_1]\times[-\phi_2,\phi_2]\), proving proposition \ref{underreac}.

\paragraph{Proof of proposition \ref{firstbest1}}
Let \(\hat{\sigma}=(\hat\sigma_m,\hat\sigma_x,\hat\sigma_r)\in\hat{\Sigma}\). Let us show that conditions (i), (ii), and (iii) are necessarily satisfied by \(\hat{\sigma}\)  -- starting from condition (iii) and proceeding backwards. 

First, note that property (iii) is equivalent to requiring \(\hat{\sigma}_r=\sigma^{\star}_r\) on path. Assume that there exist \((\omega_1,\omega_2,x)\in [-\phi_1,\phi_1]\times[-\phi_2,\phi_2]\times\mathbb{R}\) such that \(x\) arises with positive probability on path given \(\hat{\sigma}\) and \(\omega_1\), and that \(\hat{\sigma}_r(\omega_1,\omega_2,x)\ne\sigma^{\star}_r(\omega_1,\omega_2,x)\). Let \(\sigma'=(\hat\sigma_m,\hat\sigma_x,\sigma^{\star}_r)\). By definition of \(\sigma^\star_r\) and equation \ref{sstar}, \(L_{CB}(x',\sigma_r^{\star}(\omega_1',\omega_2',x'),\omega')\ge L_{CB}(x',\hat\sigma_r(\omega_1',\omega_2',x'),\omega')\) for each \((\omega_1',\omega_2',x')\in [-\phi_1,\phi_1]\times[-\phi_2,\phi_2]\times\mathbb{R}\), with \(L_{CB}(x,\sigma_r^{\star}(\omega_1,\omega_2,x),\omega)>L_{CB}(x,\hat\sigma_r(\omega_1,\omega_2,x),\omega)\) by construction. Since \((\omega_1,\omega_2,x)\) arises with positive probability given \((\hat\sigma_m,\hat\sigma_x)\) it must be that \(W(\sigma')>W(\hat\sigma)\), which contradicts \(\hat\sigma\in\hat\Sigma\). This proves (iii) by contradiction. 

For (ii), let \(m\in\mathrm{supp}\hat\sigma_m(\omega_1)\) for some \(\omega_1\in[-\phi_1,\phi_1]\). Knowing from (iii) that \(\hat\sigma_r = \sigma^{\star}_r\), and using \(g^{\prime\prime}=0\) we have that 
\begin{align*}
\mathbb{E}[L_{CB}(x,\sigma^{\star}_r(\omega_1,\omega_2,x),\omega)|m,\hat{\sigma}_m]  &= \frac{1}{2}(1-\alpha)\left[\frac{\alpha k(x)}{(1-\alpha)+\alpha k(x)}\right]^2\mathbb{E}[(\omega-\bar{x})^2|m,\hat{\sigma}_m] \\
&+ \frac{1}{2}\alpha \mathbb{E}\left[g(\bar{S}(x,\sigma^{\star}_r(\omega_1,\omega_2,x))|m,\hat{\sigma}_m\right]
 \end{align*}
where \(k:X\to\mathbb{R}\), \(k(x) = g^\prime(\bar{S}(x,r))\). Note that the above expression is convex in \(x_i\), and therefore its unique minimizer \(x\in\mathbb{R}^N\) is obtained by imposing the FOC for each \(i\in I\). This yields the following set of FOCs, for \(w_i\in[0,1]\) being the weight assigned to \(i\) in computing aggregate readjustments,
\begin{align*}
(1-\alpha)\left[\frac{\alpha k(x)}{(1-\alpha)+\alpha k(x)}\right]^2&(\mathbb{E}[\omega_1|m,\hat{\sigma}_m]-\bar{x})= \alpha k(x)\Bigg[x_i-\frac{1-\alpha}{(1-\alpha)+\alpha k(x)}\mathbb{E}[\omega_1|m,\hat{\sigma}_m] - \\
&\frac{\alpha k(x)}{(1-\alpha)+\alpha k(x)}\bar{x}\Bigg]\left(1-\frac{\alpha k(x)}{(1-\alpha)+\alpha k(x)}w_i\right),\quad \forall i\in I
\end{align*}
which is solved by \(x_i=\mathbb{E}[\omega_1|m,\hat{\sigma}_m]\) for each \(i\in I\). But this means that \(\sigma_{x_i}(m) =\mathbb{E}[\omega_1|m,\hat{\sigma}_m]\) must be true of all profiles in \(\hat\Sigma\), regardless of \(\hat{\sigma}_{x}\). This proves (ii). 

Finally, to see why it must be that \(\hat{\sigma}_m\) fully reveals \(\omega_1\), note for \(\hat{\sigma}_x,\hat{\sigma}_r\) satisfying properties (ii) and (iii), and given \(g^{\prime\prime}= 0\), \(\mathbb{E}[L_{CB}|\sigma_m\hat{\sigma}_x,\hat{\sigma}_r]\) is increasing in the residual variance \(\hat{\sigma}^2_{1}\) of \(\omega_1\) induced by the communication strategy \(\sigma_m\). Hence it must be that \(\hat{\sigma}_m\) is fully revealing.

To prove that in the game described in section \(3\) properties (i), (ii), (iii) are sufficient for a strategy profile to belong to \(\hat{\Sigma}\), it is sufficient to notice that any profile satisfying the three properties leads to the same ex-ante payoff \(\hat{W}= -\frac{1}{2}\alpha(1-\alpha)\sigma_2\).

The efficiency of the competitive equilibrium is obtained by letting \(N\to\infty\) in the equilibrium strategy profile \(\sigma^{oli}\) presented in proposition \ref{olidist} and proposition \ref{communication}. Specifically, note that \(\sigma^{oli}_r = \sigma^*_r\) so that (iii) is satisfied. The bias term in proposition \ref{olidist}.i goes to \(0\) as \(N\to\infty\) proving that unbiasedness of \(\sigma^{oli}_{x_i}\) holds to the limit, satisfying (ii). Finally note that the expression for \(\bar{P}\) in proposition \ref{communication}.ii goes to \(\infty\) as \(N\to\infty\), making equilibrium communication \(\sigma^{oli}_m\) fully informative, and satisfying (i) to the limit.

\paragraph{Proof of proposition \ref{olidist}}
We start from part (i). Fix a communication rule \(\sigma_m\in\Sigma_m\). We now derive the oligopolistic best responses \(\sigma^{oli}_{x_i}\) and \(\sigma^{oli}_r\), where \(\sigma^{oli}_r\) is a best response to \((\sigma^{oli}_{x},\sigma_m)\) and, for each \(i\in I\), \(\sigma^{oli}_{x_i}\) is a best response to \((\sigma_m, \sigma^{oli}_r)\).

First, note that it must be that \(\sigma^{oli}_r = \sigma^{\star}_r\) holds pointwise, where \(\sigma^{\star}_r\) satisfies equation \ref{sstar}. Setting \( \tilde{x}_i \equiv x_i\) in \ref{sstar} by definition, simple algebra yields 
\begin{equation}\label{sstar1}
\sigma^\star_r(\omega_1,\omega_2,x)=(1-\alpha)\omega + \alpha\bar{x}    
\end{equation}
for each \((\omega_1,\omega_2,x)\in [-\phi_1,\phi_1]\times[-\phi_2,\phi_2]\times X^I\), where \(\bar{x}=\frac{1}{N}\sum_{i\in I}x_i\).

Plugging \ref{sstar1} in \ref{payoff1}, the payoff of \(i\) as a function of \((\omega_1,\omega_2,x)\in [-\phi_1,\phi_1]\times[-\phi_2,\phi_2]\times X^I\) becomes
\begin{equation*}
u_i(\sigma^\star_r(\omega_1,\omega_2,x),x_i) = -\frac{1}{2}\left[x_i - (1-\alpha)\omega - \alpha\bar{x}\right]^2 -\beta(1-\alpha)\omega -\alpha\beta\bar{x} .
\end{equation*}
At \(t=1\) and after any message \(m\in M\), each \(i\) chooses a \(\sigma^{oli}_{x_i}(m)\) supported on the set of maximizers of \(\mathbb{E}[u_i(\sigma^\star_r(\omega_1,\omega_2,x),x_i)|m,\sigma_{x_{-i}}(m)]\) where the expectation is based on a posterior belief on \(\omega_1\) that satisfies Bayes rule whenever possible, and on the strategy profile \(\sigma_{x_{-i}}\) played by all other investors. As the objective function is strictly concave, the unique minimizer \(x_i^{\star}\) satisfies the FOC,
\begin{equation}\label{FOC1}
\left[x^{\star}_i - (1-\alpha)\mathbb{E}[\omega_1|m,\sigma_m] - \alpha\bar{x}^e\right]\left(1-\frac{\alpha}{N}\right) + \frac{\alpha\beta}{N} = 0
\end{equation}
where \(\bar{x}^e = \frac{x^{\star}_i}{N}+\sigma_{x_{-i}}(x'_{-i}|m)\sum_{j\in I\setminus\{i\}}\frac{x'_j}{N}\). First, note that the uniqueness of the minimizer implies that no mixed strategies are played. Second, the fact that \(x^{\star}_i\) depends on \(x_{j},j\ne i,\) only through the average investment \(\bar{x}\) implies symmetry of investment whenever \(m\) sent with positive probability from \(\sigma_m\).\footnote{For \(m\) sent with positive probability form \(\sigma_m\) the expectation term in equation (\ref{FOC1}) takes the same value for all investors, but after a surprising \(m\) the expectation term might differ across investors since Bayes rule does not apply.} Hence, we can let \(\bar{x}^e = x_i^{\star}\) in the condition above to obtain that the optimal strategy of each \(i\in I\) after message \(m\) occurring with positive probability given \(\sigma_m\). In particular, the profile best responses \(\sigma^{oli}_x\) must satisfy
\begin{equation}\label{invbias}
\sigma^{oli}_{x_i}(m) -  \mathbb{E}[\omega_1|m,\sigma_m] = -\frac{\alpha\beta}{(N-\alpha)(1-\alpha)}
\end{equation}
with proves part (i). 

For part (ii), let \(\sigma_m = \sigma^{com}_m\) for some \(\sigma^{com}_m\) fully revealing the state and let \(\sigma^{oli}_x\) satisfy condition (\ref{invbias}) with respect to \(\sigma^{com}_m\). On the path of play we have that for each \(\omega_1\in[-\phi_1,\phi_1]\), \(m\in\mathrm{supp}\sigma^{tr}_m(\omega_1)\) and each \(i\in I\), \(\sigma^{oli}_{x_i}(m) = \omega_1 - \frac{\alpha\beta}{(N-\alpha)(1-\alpha)}\), so that
\begin{equation}\label{barx}
\bar{x} = \omega_1 - \frac{\alpha\beta}{(N-\alpha)(1-\alpha)}.
\end{equation}
By plugging \ref{barx} in \ref{sstar1} we obtain the expression of \(r^{\star}(\sigma^{oli,tr},\omega_1,\omega_2)\), for \(\sigma^{oli,tr} = (\sigma^{com}_m,\sigma^{oli}_{x},\sigma^{\star}_r)\),
\begin{equation*}
r^{\star}(\sigma^{oli,tr},\omega_1,\omega_2) = \omega_1 + (1-\alpha)\omega_2 - \frac{\alpha^2\beta}{(N-\alpha)(1-\alpha)},
\end{equation*}
which proves part (ii).

\paragraph{Proof of proposition \ref{communication}} First, note that in any PBE it must be that \(CB\) optimally plays the strategy given by (\ref{sstar1}) a \(t = 2\). Fixing this rate rule, we can write the expected loss of \(CB\) given state \(\omega_1\in[-\phi_1,\phi_1]\) as a function of \(x\in X^I\) is
\begin{equation}\label{CSderiv}
\mathbb{E}[L_{CB}(x,\omega)|\omega_1] = \frac{1}{2}\alpha^2(1-\alpha)\mathbb{E}[(\bar{x}-\omega)^2|\omega_1] + \alpha\frac{1}{2N}\sum_{i\in I}\mathbb{E}\left[[(1-\alpha)\omega + \alpha\bar{x} - x_i]^2|\omega_1\right]
\end{equation}
As shown in the previous proof, any equilibrium will induce \(x_i=x_j\) for each \(i,j\in I\) regardless of the communication strategy employed, so communication in this game cannot be used, in any an equilibrium, to induce heterogeneity in investment choices.\footnote{Per proposition \ref{firstbest1}.ii, \(CB\) would not benefit from heterogeneous investment plans \(x\in X^I\), provided that plans are unbiased.} Therefore, we restrict the attention the profiles in the set \(\bar{X} = \left\{x\in X:x_i = x_j\quad \forall i,j\in I\right\}\). For \(x\in \bar{X}\), (\ref{CSderiv}) simplifies to
\begin{align*}
\mathbb{E}[L_{CB}(x,\omega)|\omega_1] &= \frac{1}{2}\alpha(1-\alpha)\mathbb{E}[(\bar{x}-\omega)^2|\omega_1] \\
& = \frac{1}{2}\alpha(1-\alpha)(\bar{x} - \omega_1)^2 + \frac{1}{2}\alpha(1-\alpha)\sigma_2^2.
\end{align*}
It is immediate to see from the above expression, that, in any PBE, the \(CB\) problem simplifies to choosing some \(\sigma^{oli}_m\in\Sigma_m\) that minimizes \(\mathbb{E}[(\bar{x} - \omega_1)^2]\) \textit{given} the PBE strategies \((\sigma^{oli}_x,\sigma^{\star}_r)\). But we know from (\ref{invbias}) that, for each \(i\in I\),
\begin{equation*}
\sigma^{oli}_{x_i}(m) = \mathbb{E}[\omega_1|m,\sigma^{oli}_m] -\frac{\alpha\beta}{(N-\alpha)(1-\alpha)}
\end{equation*}
must be satisfied in equilibrium. Focusing on communication strategies \(\sigma_m\) for which each message occurs with positive probability in at least one state, the set of PBE communication strategies \(\sigma^{oli}_m\) is the argmax of the following problem
\begin{align} \label{optimization}
\min\limits_{\sigma_m\in\Sigma_m}& \int_{-\phi_1}^{\phi_1}\int_{-\phi_1}^{\phi_1}(\sigma_{x_i}^{oli}(m) - \omega_1)^{2}\sigma_m(m|\omega_1)(2\phi_1)^{-1}d m d\omega_1 \\ 
& \text{ s.t. } \quad \text{(i) }\sigma^{oli}_{x_i}(m) = \mathbb{E}[\omega_1|m,\sigma_m]  -\frac{\alpha\beta}{(N-\alpha)(1-\alpha)},  \notag \\
&\quad\quad\quad \text{(ii) }\forall m\in M, \exists \omega_1\in [-\phi_1,\phi_1] : \sigma_m(m|\omega_1)>0. \notag
\end{align}

The general solution of this type of problem has been characterized by \citet{CS}, hereafter referred to as CS. Note that relaxing (ii) would not deliver economically different communication strategies/equilibria provided that after messages \(m\) that are unexpected (i.e., with \(\sigma_m(m|\omega_1)=0\) for each \(\omega_1\in[-\phi_1,\phi_1]\)), we restrict off-path beliefs to be such that for each \(i\in I\) the best response to \(m\) does not expand the set of \(x_i\) played on the equilibrium path.

The mapping to CS is particularly evident by noticing that we can transform the variables and parameters of interest in our setting to match a specific case of their formulation. First, define the following transformation: \(t: \mathbb{R}\to \mathbb{R}, t(x) = (2\phi_1)^{-1}\left(x+\phi_1+\frac{\alpha\beta}{(N-\alpha)(1-\alpha)}\right)\). For each \(\omega_1\in[-\phi_1,\phi_1]\) we can apply the transformation \(t\) to the perfect information ideal investment levels of \(CB\) and the representative investor \(i\), denoted as \(x^{\star}_{CB}(\omega_1) = \omega_1\) and \(x^{\star}_i(\omega_1)=\omega_1 - \frac{\alpha\beta}{(N-\alpha)(1-\alpha)}\) respectively. It is easy to see that \(t(x^{\star}_{i}(\omega_1)) = (2\phi_1)^{-1}(\omega_1+\phi_1)\) and \(t(x^{\star}_{CB}(\omega_1)) = t(x^{\star}_{i}(\omega_1)) +b\) for \(b = \frac{\alpha\beta}{2\phi_1(N-\alpha)(1-\alpha)}\). Denote by \(t^\star_{\omega_1}\) the random variable equal to \((2\phi_1)^{-1}(\omega_1+\phi_1)\), the transformed ideal point of \(i\). First, note that \(t^{\star}_{\omega_1}\sim U[0,1]\), supported on the unit interval as in CS. Second, note that for each \(\omega_1\in [-\phi_1,\phi_1]\) and \(x_i\in\mathbb{R}\) it has
\begin{equation*}
\left(\omega_1 - \frac{\alpha\beta}{(N-\alpha)(1-\alpha)} - x_i\right) \propto (t^\star_{\omega_1} - t(x_i))
\end{equation*}
and, similarly, 
\begin{equation*}
(\omega_1 - x_i) \propto  (t_{\omega_{1}}+b - t(x_i))
\end{equation*}
Third, note that \(t\) is a bijection between \(X\) and \(X\) and is therefore invertible. 

Substantially, \(t\) creates a``bridge" between our problem (\ref{optimization}) and the seminal example studied by CS (section 4). Consider that example, and set \(b = \frac{\alpha\beta}{2\phi_1(N-\alpha)(1-\alpha)}\). The above relations imply that (i) for every partition PBE of Example 4 where communication partitions \([0,1]\) in \(P\) intervals with cutoffs \(a_0<a_1<...< a_P\), there exists a PBE of our our game where \(\sigma^{oli}_m\) partitions \([-\phi_1,\phi_1]\) in \(P\) intervals, with cutoffs \(t^{-1}(a_0 +b)<t^{-1}(a_1 +b)<...<t^{-1}(a_P +b)\); and (ii) for every partition PBE of our game where \(\sigma^{oli}_m\) partitions \([-\phi_1,\phi_1]\) in \(P\) intervals with cutoffs \(a_0<a_1<...< a_P\), there exists a PBE of CS's seminal example, where communication partitions \([0,1]\) in \(P\) intervals, with cutoffs \(t(a_0)-b<t(a_1)-b<...<t(a_P)-b\). This correspondence between equilibrium communication strategies in the two games also implies that the strong comparative statics in section 4 of CS extend to our setting for \(b = \frac{\alpha\beta}{2\phi_1(N-\alpha)(1-\alpha)}\), proving proposition \ref{communication}.i-ii.

\paragraph{Proof of proposition \ref{kites}}
Let \(b:[0,1)\to\mathbb{R}, b(\tilde{\alpha})=-\frac{\tilde{\alpha}\beta}{(N-\tilde{\alpha})(1-\tilde{\alpha})}\), mapping central banker types to the corresponding investment bias in equilibrium (we ignore the case in which \(\tilde{\alpha} = 1\) as such case leads to an infinite welfare loss). By repeating the steps of the proof of proposition \ref{olidist} replacing \(\alpha\) with \(\tilde{\alpha}\in[0,1)\), it is easily shown that, for each \(m\in M\), \((\omega_1,\omega_2)\in[-\phi_1,\phi_1]\times[-\phi_2,\phi_2]\) and \(x\in X^I\),
\begin{gather*}
\sigma^{\star}_{r}(\omega_1,\omega_2,x;\tilde{\alpha}) = (1-\tilde{\alpha})\omega+\tilde{\alpha}\bar{x} \\
\sigma_{x_i}^{oli}(m;\tilde{\alpha}) = \mathbb{E}[\omega_1|m,\sigma_m] + b(\tilde{\alpha}).
\end{gather*}
It follows that the ex-ante loss for the central bank takes the following form, for \(\sigma^{oli,tr}\) and \(\sigma^{oli}_{\bar{P}}\) respectively,
\begin{align}
W(\sigma^{oli,tr}(\tilde{\alpha})) &= -\frac{1}{2}[\tilde{\alpha}^2(1-\alpha)+\alpha(1-\tilde{\alpha})^2](\sigma_2^2 +  b(\tilde{\alpha})^2) \label{wtr}\\
W(\sigma^{oli}_{\bar{P}}(\tilde{\alpha})) &= -\frac{1}{2}[\tilde{\alpha}^2(1-\alpha)+\alpha(1-\tilde{\alpha})^2](\hat{\sigma}_{1,\bar{P}}^2(\tilde{\alpha}) + \sigma_2^2 +  b(\tilde{\alpha})^2)\label{wcheap}
\end{align}
where the only difference is that in (\ref{wtr}) the residual variance after communication is zero because communication is fully informative. 

First, consider part (i) of the proposition. It is sufficient to compare \(W(\sigma^{oli,tr}(\alpha))\) and \(W(\sigma^{oli}_{\bar{P}}(\alpha))\) with \(W(\sigma^{oli,tr}(0))\) and \(W(\sigma^{oli}_{\bar{P}}(0))\) respectively, by plugging the corresponding values of \(\tilde{\alpha})\) in \ref{wtr} and \ref{wcheap}. It has 

\begin{align*}
W(\sigma^{oli,tr}(0))> W(\sigma^{oli,tr}(\alpha)) & \iff \sigma_2 < \sqrt{\frac{\alpha}{1-\alpha}}\left(\frac{\beta}{N-\alpha}\right) \equiv \sigma_2^{oli,tr} \\
W(\sigma^{oli}(0))> W(\sigma^{oli}(\alpha)) & \iff \sigma_2 < \sqrt{\frac{\alpha}{1-\alpha}\left(\frac{\beta}{N-\alpha}\right)^2 + \frac{1-\alpha}{\alpha}\hat{\sigma}_{1,\bar{P}}^2(\tilde{\alpha})} \equiv \sigma_2^{oli},
\end{align*}
and indeed it holds \(\sigma_2^{oli,tr} <\sigma_2^{oli}\) because \(\hat{\sigma}_{1,\bar{P}}^2(\tilde{\alpha})>0\), which proves part (i).

Next, we prove part (ii). For the second statement, note that \(W(\sigma^{oli,tr}(\tilde{\alpha}))\ge W(\sigma^{oli}(\tilde{\alpha}))\) for all \(\tilde{\alpha}\in(0,1)\). Hence it must be \(W(\sigma^{oli,tr}(\tilde{\alpha}^{oli}))\ge W(\sigma^{oli}(\tilde{\alpha}^{oli}))\). But by definition  \(W(\sigma^{oli,tr}(\tilde{\alpha}^{oli,tr}))\ge W(\sigma^{oli,tr}(\tilde{\alpha}^{oli}))\), implying \(W(\sigma^{oli,tr}(\tilde{\alpha}^{oli,tr}))\ge W(\sigma^{oli}(\tilde{\alpha}^{oli}))\).

We now turn to the first statement of part (ii). First, let us take the derivative of (\ref{wtr}) with respect to \(\tilde{\alpha}\), yielding,
\begin{equation*}
\frac{\partial W(\sigma^{oli,tr}(\tilde{\alpha}))}{\partial\tilde{\alpha}} = (\alpha-\tilde{\alpha})(\sigma_2^2 +  b(\tilde{\alpha})^2) - [\tilde{\alpha}^2(1-\alpha)+\alpha(1-\tilde{\alpha})^2]\underbrace{\frac{\partial b(\tilde{\alpha})}{\partial \tilde{\alpha}}b(\tilde{\alpha})}_{>0}.
\end{equation*}
On the one hand, the above expression is always negative for \(\tilde{\alpha} \ge \alpha\), implying that the optimal central banker, if it exists, must be kitish. On the other hand, the derivative is positive for \(\tilde{\alpha} = 0\), implying that \(\tilde{\alpha}^{oli,tr}\in(0,\alpha)\), is the optimal central banker exists. Existence follows from the fact that \([0,\alpha]\) is compact and (\ref{wtr}) continuous. 

From the first order condition, we obtain following equality,

\begin{equation}\label{plug1}
\sigma_2^2 +  b(\tilde{\alpha})^2 =  \left[\frac{\tilde{\alpha}^2(1-\alpha)+\alpha(1-\tilde{\alpha})^2}{\alpha-\tilde{\alpha}}\right]\frac{\partial b(\tilde{\alpha})}{\partial \tilde{\alpha}}b(\tilde{\alpha})     
\end{equation}

Note that 
\begin{align*}
   \frac{\partial b(\tilde{\alpha})}{\partial \tilde{\alpha}} &= -\frac{\beta(N-\tilde{\alpha})(1-\tilde{\alpha}) + \tilde{\alpha}\beta(1+N-2\tilde{\alpha})}{(1-\tilde{\alpha})^2(N-\tilde{\alpha})^2}=-\frac{(N-\tilde{\alpha}^2)\beta}{(1-\tilde{\alpha})^2(N-\tilde{\alpha})^2} \\
   \frac{\partial^2 b(\tilde{\alpha})}{\partial \tilde{\alpha}^2} &=2 \frac{\tilde{\alpha}\beta(1-\tilde{\alpha})(N-\tilde{\alpha})-(N-\tilde{\alpha}^2)\beta(1+N-2\tilde{\alpha})}{(1-\tilde{\alpha})^3(N-\tilde{\alpha})^3}\\
   &=2\frac{\tilde{\alpha}\beta N -\tilde{\alpha}^2\beta N-\tilde{\alpha}^2\beta +\tilde{\alpha}^3\beta -N\beta+\tilde{\alpha}^2\beta-N^2\beta+\tilde{\alpha}^2\beta N+2\tilde{\alpha}\beta N-2\tilde{\alpha}^3\beta}{(1-\tilde{\alpha})^3(N-\tilde{\alpha})^3}\\
   &=2\beta\frac{3\tilde{\alpha} N -N-N^2 -\tilde{\alpha}^3}{(1-\tilde{\alpha})^3(N-\tilde{\alpha})^3}.
\end{align*}
We want to show that \( W(\sigma^{oli,tr}(\tilde{\alpha}))\) is concave in \(\tilde{\alpha}\). It is easy to see that
\begin{align*}
\frac{\partial^2 W(\sigma^{oli,tr}(\tilde{\alpha}))}{\partial\tilde{\alpha}^2}&= -(\sigma_2^2 +  b(\tilde{\alpha})^2)+4(\alpha-\tilde{\alpha})\frac{\partial b(\tilde{\alpha})}{\partial \tilde{\alpha}}b(\tilde{\alpha}) + \\
&\quad - [\tilde{\alpha}^2(1-\alpha)+\alpha(1-\tilde{\alpha})^2] \left[\frac{\partial^2 b(\tilde{\alpha})}{\partial \tilde{\alpha}^2}b(\tilde{\alpha})+\left(\frac{\partial b(\tilde{\alpha})}{\partial \tilde{\alpha}}\right)^2\right].
\end{align*}
Using \ref{plug1} we can rewrite the above expression as 
\begin{align*}
\frac{\partial^2 W(\sigma^{oli,tr}(\tilde{\alpha}))}{\partial\tilde{\alpha}^2}&= \frac{4(\alpha-\tilde{\alpha})^2 - [\tilde{\alpha}^2(1-\alpha)+\alpha(1-\tilde{\alpha})^2]}{\alpha-\tilde{\alpha}}\frac{\partial b(\tilde{\alpha})}{\partial \tilde{\alpha}}b(\tilde{\alpha}) + \\
&\quad - [\tilde{\alpha}^2(1-\alpha)+\alpha(1-\tilde{\alpha})^2] \left[\frac{\partial^2 b(\tilde{\alpha})}{\partial \tilde{\alpha}^2}b(\tilde{\alpha})+\left(\frac{\partial b(\tilde{\alpha})}{\partial \tilde{\alpha}}\right)^2\right] \\
&= \frac{4(\alpha-\tilde{\alpha})^2 - [\tilde{\alpha}^2(1-\alpha)+\alpha(1-\tilde{\alpha})^2]}{\alpha-\tilde{\alpha}}\frac{\partial b(\tilde{\alpha})}{\partial \tilde{\alpha}}b(\tilde{\alpha}) + \\
&\quad - \frac{[\tilde{\alpha}^2(1-\alpha)+\alpha(1-\tilde{\alpha})^2](\alpha-\tilde{\alpha})}{(\alpha-\tilde{\alpha})} \left[\frac{\partial^2 b(\tilde{\alpha})}{\partial \tilde{\alpha}^2}\left(\frac{\partial b(\tilde{\alpha})}{\partial \tilde{\alpha}}\right)^{-1} + \right. \\
&\quad \left. + \left(\frac{\partial b(\tilde{\alpha})}{\partial \tilde{\alpha}}\right)b(\tilde{\alpha})^{-1}\right]\frac{\partial b(\tilde{\alpha})}{\partial \tilde{\alpha}}b(\tilde{\alpha}),
\end{align*}
and that we know that at the optimum \(\alpha-\tilde{\alpha}>0\) so that the sign of the derivative is the same of the sign of
\begin{align*}
4(\alpha-\tilde{\alpha})^2 - \left[\tilde{\alpha}^2(1-\alpha)+\alpha(1-\tilde{\alpha})^2\right]\left\{1 + (\alpha-\tilde{\alpha})\left[\frac{\partial^2 b(\tilde{\alpha})}{\partial \tilde{\alpha}^2}\left(\frac{\partial b(\tilde{\alpha})}{\partial \tilde{\alpha}}\right)^{-1}+\left(\frac{\partial b(\tilde{\alpha})}{\partial \tilde{\alpha}}\right)b(\tilde{\alpha})^{-1}\right]\right\}. 
\end{align*}
Note that \(\frac{\partial^2 b(\tilde{\alpha})}{\partial \tilde{\alpha}^2}\left(\frac{\partial b(\tilde{\alpha})}{\partial \tilde{\alpha}}\right)^{-1} > 2\frac{(1+N-2\tilde{\alpha})-\tilde{\alpha}(1-\tilde{\alpha})}{(1-\tilde{\alpha})(N-\tilde{\alpha})}>\frac{2}{1-\tilde{\alpha}}\), and that \(\left(\frac{\partial b(\tilde{\alpha})}{\partial \tilde{\alpha}}\right)b(\tilde{\alpha})^{-1} = \frac{N-\tilde{\alpha}^2}{\alpha(N-\tilde{\alpha})(1-\alpha)}>\frac{1}{\tilde{\alpha}(1-\tilde{\alpha})}\). So that it holds that 
\begin{equation*}
4(\alpha-\tilde{\alpha})^2 - \left[\tilde{\alpha}^2(1-\alpha)+\alpha(1-\tilde{\alpha})^2\right]\left[1 + (\alpha-\tilde{\alpha})\frac{1+2\tilde{\alpha}}{\tilde{\alpha}(1-\tilde{\alpha})}\right]\le 0\implies \frac{\partial^2 W(\sigma^{oli,tr}(\tilde{\alpha}))}{\partial\tilde{\alpha}^2} < 0.   
\end{equation*}
Standard algebra shows that the expression on the LHS above has the same sign as
\begin{equation*}
h(\tilde{\alpha},\alpha) = 4\alpha^2\tilde{\alpha} + 4\tilde{\alpha}^3-6\alpha\tilde{\alpha}^2-\tilde{\alpha}^4-\alpha^2.
\end{equation*}
when \(\tilde{\alpha}\) is at the optimum. Fix \(\alpha\in(0,1)\). We start by showing that \(h(\tilde{\alpha},\alpha)\) is monotone in \(\tilde{\alpha}\). First, note that \(h_1(\tilde{\alpha},\alpha)\ge0 \iff \alpha^2 + 3\tilde{\alpha}^2-3\alpha\tilde{\alpha}-\tilde{\alpha}^3\ge0\). Second, note that \(\alpha^2 + 3\tilde{\alpha}^2-3\alpha\tilde{\alpha}-\tilde{\alpha}^3\) is decreasing in \(\tilde{\alpha}\) on \([0,1-\sqrt{1-\alpha})\) and increasing in \(\tilde{\alpha}\) on \((1-\sqrt{1-\alpha},1]\), while it achieves a minimum at \(\tilde{\alpha}_0 = 1-\sqrt{1-\alpha}\). By substitution one can easily verify that \(\alpha^2 + 3\tilde{\alpha}_0^2-3\alpha\tilde{\alpha}_0-\tilde{\alpha}_0^3\ge 0 \iff 2(1-\sqrt{1-\alpha})-\alpha\ge 0\). But note that \(2(1-\sqrt{1-\alpha})-\alpha\ge 0\) is increasing in \(\alpha\) and it is non-negative for both \(\alpha = 0\) and \(\alpha = 1\), from which follows that \(2(1-\sqrt{1-\alpha})-\alpha\ge 0\), and hence \(h_1(\tilde{\alpha},\alpha)\ge0\). Given the monotonicity of \(h_1\), it is sufficient to prove that \(h(0,\alpha)\) and \(h(\alpha,\alpha)\) are both negative. This is easily verified, given that \(h(0,\alpha) = -\alpha<0\) and \(h(\alpha,\alpha) = -(1-\alpha)^2<0\). This implies that, whenever \(\alpha\in (0,1)\), at the optimum it holds \(\frac{\partial^2 W(\sigma^{oli,tr}(\tilde{\alpha}))}{\partial\tilde{\alpha}^2} < 0\).

We use the concavity result to establish the comparative statics of proposition \ref{kites} (ii) using the implicit function theorem. First, note that \(\frac{\partial^2 W(\sigma^{oli,tr}(\tilde{\alpha}))}{\partial\tilde{\alpha}\partial \sigma_2^2}=\alpha-\tilde{\alpha}>0\) implies \(\frac{\partial\tilde{\alpha}^{oli,tr}}{\partial \sigma_2^2}>0\). Second, to see that the optimal central banker is increasingly \textit{quailish} the more competitive the banking sector, note that 
\begin{equation*}
\frac{\partial W(\sigma^{oli,tr}(\tilde{\alpha}))}{\partial\tilde{\alpha}\partial N}=-2(\alpha-\tilde{\alpha})\frac{(1-\tilde{\alpha})(\tilde{\alpha}\beta)^2}{(1-\tilde{\alpha})^3(N-\tilde{\alpha})^3}+\left[\tilde{\alpha}^2(1-\alpha)+\alpha(1-\tilde{\alpha})^2\right]\frac{\tilde{\alpha}\beta^2\left(3\frac{N-\tilde{\alpha}^2}{N-\tilde{\alpha}}-1\right)}{(1-\tilde{\alpha})^3(N-\alpha)^3}
\end{equation*}
so that 
\begin{equation*}
    \frac{\partial W(\sigma^{oli,tr}(\tilde{\alpha}))}{\partial\tilde{\alpha}\partial N} > 0 \iff -2(\alpha-\tilde{\alpha})\tilde{\alpha} + \left[\tilde{\alpha}^2\frac{1-\alpha}{1-\tilde{\alpha}}+\alpha(1-\tilde{\alpha})\right]\left(3\frac{N-\tilde{\alpha}^2}{N-\tilde{\alpha}}-1\right)>0
\end{equation*}
But \(-2(\alpha-\tilde{\alpha})\tilde{\alpha} + \left[\tilde{\alpha}^2\frac{1-\alpha}{1-\tilde{\alpha}}+\alpha(1-\tilde{\alpha})\right]\left(3\frac{N-\tilde{\alpha}^2}{N-\tilde{\alpha}}-1\right)> 2\left[\alpha(1-\tilde{\alpha}) - \tilde{\alpha}(\alpha-\tilde{\alpha})\right]>0\), since \(0<\tilde{\alpha}<\alpha<1\). Hence, we have that \(\frac{\partial W(\sigma^{oli,tr}(\tilde{\alpha}))}{\partial\tilde{\alpha}\partial N} > 0\), implying \(\frac{\partial\tilde{\alpha}^{oli,tr}}{\partial N}>0\). To see that \(\frac{\partial\tilde{\alpha}^{oli,tr}}{\partial \alpha}>0\), it is sufficient to prove that \(\frac{\partial^2 W(\sigma^{oli,tr}(\tilde{\alpha}))}{\partial\tilde{\alpha}\partial \alpha}<0\). Note that
\begin{align*}
 \frac{\partial^2 W(\sigma^{oli,tr}(\tilde{\alpha}))}{\partial\tilde{\alpha}\partial \alpha} &= (\sigma_2^2+b(\tilde{\alpha})^2)- (1-2\tilde{\alpha})\frac{\partial b(\tilde{\alpha})}{\partial \tilde{\alpha}}b(\tilde{\alpha}) \\
 &= \left\{\left[\frac{\tilde{\alpha}^2(1-\alpha)+\alpha(1-\tilde{\alpha})^2}{\alpha-\tilde{\alpha}}\right] - (1-2\tilde{\alpha})\right\}\frac{\partial b(\tilde{\alpha})}{\partial \tilde{\alpha}}b(\tilde{\alpha}) \\
 & = \left[\frac{\tilde{\alpha}(1-\tilde{\alpha})}{\alpha-\tilde{\alpha}}\right]\frac{\partial b(\tilde{\alpha})}{\partial \tilde{\alpha}}b(\tilde{\alpha})>0,
\end{align*}
so that \(\frac{\partial\tilde{\alpha}^{oli,tr}}{\partial \alpha}>0\) follows from the implicit function theorem.

The main result for \(\tilde{\alpha}^{oli}\) is derived analogously, from (\ref{wcheap}). Note that
\begin{align*}
W(\sigma^{oli}_{\bar{P}}(\tilde{\alpha})) &= W(\sigma^{oli,tr}_{\bar{P}}(\tilde{\alpha})) - \frac{1}{2}[\tilde{\alpha}^2(1-\alpha)+\alpha(1-\tilde{\alpha})^2]\hat{\sigma}_{1,\bar{P}}^2(\tilde{\alpha})\label{wcheap}.
\end{align*}
The existence of \(\tilde{\alpha}^{oli}\) is guaranteed by the continuity of \(\hat{\sigma}_{1,\bar{P}}^2(\tilde{\alpha})\). For \(\tilde{\alpha}^{oli}\in(0,\alpha)\) to hold, it is sufficient to show that 
\begin{equation*}
- \frac{1}{2}[\tilde{\alpha}^2(1-\alpha)+\alpha(1-\tilde{\alpha})^2]\hat{\sigma}_{1,\bar{P}}^2(\tilde{\alpha})
\end{equation*}
is weakly decreasing in \(\tilde{\alpha}\) if \(\tilde{\alpha}\ge \alpha\) and weakly increasing in \(\tilde{\alpha}\) in a (right) neighborhood of \(\tilde{\alpha} = 0\). 
First, given that \(\hat\sigma_{1\bar{P}}(\tilde{\alpha})\) is always weakly increasing in \(\tilde{\alpha}\) and \(\tilde{\alpha}^2(1-\alpha)+\alpha(1-\tilde{\alpha})^2\) is weakly increasing in \(\tilde{\alpha}\) if \(\tilde{\alpha}\ge \alpha\), we have that \(\tilde{\alpha} < \alpha\). Second, the right derivative of \(\hat\sigma_{1\bar{P}}(\tilde{\alpha})\) at \(\alpha=0\) is equal to 0, which guarantees \(\tilde{\alpha}^{oli}>0\). 

Finally, we prove part (iii). First, to see that under transparent communication markets are worse off than under transparency, note that market players' payoff as a function of \(\tilde{\alpha}\) are 
\begin{align*}
EU^{oli,tr}_{i} =& -\frac{1}{2}(1-\tilde{\alpha})^2(\sigma^2_2 + \beta(\tilde{\alpha})^2) + \frac{(\tilde{\alpha}\beta)^2}{(N-\tilde{\alpha})(1-\tilde{\alpha})} \\
=& -\frac{1}{2}(1-\tilde{\alpha})^2\sigma^2_2 + \frac{(2N-1-\tilde{\alpha})(\tilde{\alpha}\beta)^2}{2(N-\tilde{\alpha})^2(1-\tilde{\alpha})} \\
=& -\frac{1}{2}(1-\tilde{\alpha})^2\sigma^2_2 + \left(\frac{\tilde{\alpha}\beta}{N-\tilde{\alpha}}\right)^2\left(\frac{N-1}{1-\tilde{\alpha}} + \frac{1}{2}\right)
\end{align*}
which is clearly increasing in \(\tilde{\alpha}\).
To see that under cheap talk communication markets can be better off with a kitish central banker relative to an unbiased one, it is sufficient to find an example. The example is provided at the end of section \(\ref{sectkites}\).

\paragraph{Proof of proposition \ref{dynfirstbest}}
We want to show that there exist a \(\delta^\star\in (0,1)\) such that the proposed strategy is a PBE of the repeated game with discount rate \(\delta\ge\delta^\star\). Note that in the punishment phase players revert to a fixed stage game PBE, so that it is sufficient to check that, on the path of play, no player has incentive to deviate. 

Consider \(CB\) first. Note that \(CB\) actions at time \(s<\tau\) do not influence period \(\tau\) play of any player, so it is sufficient to check that \(CB\) has no way to increase its stage payoff via a deviation. The latter statement follows immediately from the fact that stage game play on the path of play satisfy the efficiency conditions of \ref{firstbest1}. 

Turning to the deviation incentives of player \(i\in I\), the no profitable deviation condition requires

\begin{equation*}
 \frac{1}{2}\left[\frac{\alpha\beta}{N-\alpha}\right]^2 + \frac{\delta}{1-\delta}\left\{\frac{1}{2}\left[\frac{\alpha\beta}{N-\alpha}\right]^2\frac{2N-1-\alpha}{(1-\alpha)} -\frac{1}{2}(1-\alpha)^2 Var(\omega_1) \right\} \le 0,
\end{equation*}
where the left hand side is the net benefit that \(i\) obtains if she deviates to the stage best response once and then follows the equilibrium strategy in all future stages. Using \(Var(\omega_1) = \frac{\phi_1^2}{3}\), it is easily seen from the above condition that
\begin{equation*}
\delta^\star \equiv \frac{\left[\frac{\alpha\beta}{N-\alpha}\right]^2}{\left[\frac{\alpha\beta}{N-\alpha}\right]^2 \left[1 - \frac{2N-1-\alpha}{1-\alpha} \right] + \frac{\phi_1^2}{3}(1-\alpha)^2}.
\end{equation*}
For the equilibrium to exist it must be that \(\delta^\star<1\), which is true if and only if \(\phi_1>\frac{\alpha\beta}{(N-\alpha)(1-\alpha)}\sqrt{\frac{2N-1-\alpha}{1-\alpha}}\). 

\paragraph{Proof of proposition \ref{coord}}
For part (i), it is sufficient to provide a collusive \(PBE\) strategy profile that implements efficient play on path in every stage, that is, such that stage play on the equilibrium path satisfies the conditions of \ref{firstbest1}. We will then verify that such equilibrium exists if \(\hat{\sigma}^2_{1,\bar{P}}(N)>\left[\frac{\alpha\beta}{(N-\alpha)(1-\alpha)}\right]^2\frac{2N-\alpha-1}{1-\alpha}\). 

Consider the following profile. At \(\tau = 0\), \(CB\) chooses \(m^\tau = \omega^\tau_1\); at \(\tau> 0\), \(CB\) select \(m^\tau = \omega^\tau_1\) if \(x^s_i = m^s\) for each \(s<\tau\) and \(i\in I\); if there exists \(s<\tau\) and \(i\in I\) such that \(x^s_i\ne m^s\), then \(CB\) uses the a stage communication rule \(\sigma_{m,\bar{P}}\) corresponding to the most informative stage game PBE when the bias is at the no-coordination \(\frac{\alpha\beta} {(N -\alpha)( 1- \alpha)}\). Each investor \(i\in I\) plays the following strategy. At \(\tau=0\), \(x_i^\tau = m^\tau\). At \(\tau> 0\), \(i\in I\) selects \(x_i^\tau = m^\tau\) if \(x^s_j = m^s\) for each \(s<\tau\) and \(j\in I\); if there exists \(s<\tau\) and \(j\in I\) such that \(x^s_j\ne m^s\), then \(x_i^\tau= \mathbb{E}[\omega^\tau_1|m^\tau,\sigma_{m,\bar{P}}] -\frac{\alpha\beta} {(N -\alpha)(1- \alpha)}\). At each \(\tau\ge 0\), \(CB\) uses the policy rule \(r^\tau = (1-\alpha)\omega^\tau+\alpha\bar{x}^\tau\). 

By construction, in every stage the \(CB\) communicates as much as possible given the equilibrium expected investment bias of that period. As in the case of the previous proposition, punishment is carried out by playing a fixed stage Nash in any period after a deviation (regardless of how the history of play evolves after the deviation). Hence deviations are not profitable during the punishment phase. Moreover, the \(CB\) has no incentive to deviate from the equilibrium path, since efficient play is implemented on the equilibrium path in each stage. Hence, for the strategy profile considered to be a PBE of the repeated game it is sufficient to impose that \(i\in I\) has no incentive to make a one shot deviation from the equilibrium path. This requires

\begin{equation*}
 \frac{1}{2}\left[\frac{\alpha\beta}{N-\alpha}\right]^2 + \frac{\delta}{1-\delta}\left\{\frac{1}{2}\left[\frac{\alpha\beta}{N-\alpha}\right]^2\frac{2N-1-\alpha}{1-\alpha} -\frac{1}{2}(1-\alpha)^2 \hat{\sigma}_{1,\bar{P}}^2(N) \right\} \le 0,
\end{equation*}
where the left hand side is the net benefit that \(i\) obtains if she does a one-shot deviation towards the stage best response and then follows the equilibrium strategy in all future stages. Note that at \(\delta = \delta^\star_1\) the previous equation must hold with equality. Simple algebra yields
\begin{equation*}
\delta^\star_1 \equiv \frac{\left[\frac{\alpha\beta}{N-\alpha}\right]^2}{\left[\frac{\alpha\beta}{N-\alpha}\right]^2 \left[1 - \frac{2N-1-\alpha}{1-\alpha} \right] + (1-\alpha)^2 \hat{\sigma}_{1,\bar{P}}^2(N)}.
\end{equation*}
For the equilibrium to exist it must be that \(\delta^\star<1\), which is true if and only if \(\hat{\sigma}_{1,\bar{P}}^2(N)>\left[\frac{\alpha\beta}{(N-\alpha)(1-\alpha)}\right]^2\frac{2N-1-\alpha}{1-\alpha}\). Note that the latter inequality also guarantees that \(i\)'s stage-game payoff under the equilibrium considered is greater than under the most informative stage-game equilibrium for the parameters considered.

For part (ii), it is sufficient to provide a collusive \(PBE\) strategy profile that mimics the stage game equilibrium when \(N=1\) on the path of play in every stage. We will then verify that the existence of such equilibrium requires \(\hat{\sigma}^2_{1,\bar{P}}(1) - \hat{\sigma}^2_{1,\bar{P}}(N)<\left[\frac{\alpha\beta}{(1-\alpha)^2}\right]^2\). 

Consider the following profile. At \(\tau = 0\), \(CB\) uses a stage communication rule \(\sigma^1_{m,\bar{P}}\) corresponding to the most informative stage game PBE when investors' bias is at the monopolistic level \(\frac{\alpha\beta} {( 1- \alpha)^2}\); at \(\tau> 0\), \(CB\) keeps using the same communication rule if \(x_i^s= \mathbb{E}[\omega^s_1|m^s,\sigma^1_{m,\bar{P}}] -\frac{\alpha\beta} {(1- \alpha)^2}\) for each \(s<\tau\) and \(i\in I\); if there exists \(s<\tau\) and \(i\in I\) such that \(x^s_i\ne m^s\), then \(CB\) uses a stage communication rule \(\sigma^N_{m,\bar{P}}\) corresponding to the most informative stage game PBE when the bias is at the no-coordination level \(\frac{\alpha\beta} {(N -\alpha)( 1- \alpha)}\). Each investor \(i\in I\) plays the following strategy. At \(\tau=0\), \(x_i^\tau= \mathbb{E}[\omega^\tau_1|m^\tau,\sigma^1_{m,\bar{P}}] -\frac{\alpha\beta} {(1- \alpha)^2}\). At \(\tau> 0\), \(i\in I\) selects \(x^\tau_i = \mathbb{E}[\omega^\tau_1|m^\tau,\sigma^1_{m,\bar{P}}] -\frac{\alpha\beta} {(1- \alpha)^2}\) if \(x^s_j = m^s\) for each \(s<\tau\) and \(j\in I\); if there exists \(s<\tau\) and \(j\in I\) such that \(x^s_j \ne \mathbb{E}[\omega^s_1|m^s,\sigma^1_{m,\bar{P}}] -\frac{\alpha\beta} {(1- \alpha)^2}\), then \(x^\tau_i = \mathbb{E}[\omega^\tau_1|m^\tau,\sigma^N_{m,\bar{P}}] -\frac{\alpha\beta} {(N-\alpha)(1- \alpha)}\). At each \(\tau\ge 0\), \(CB\) uses the policy rule \(r^\tau = (1-\alpha)\omega^\tau+\alpha\bar{x}^\tau\). 

By construction, in every stage the \(CB\) communicates as much as possible given the equilibrium expected investment bias of that period. As in the previous cases, punishment is carried out by playing a fixed stage Nash in any period after a deviation (regardless of how the history of play evolves after the deviation). Hence deviations are not profitable during the punishment phase. The \(CB\) has no incentive to deviate from the equilibrium path, since it is maximizing its stage game expected payoff given the investors' strategy, and \(CB\) deviations have no influence on subsequent play. Hence, for the strategy profile outlined to be a PBE of the repeated game it is sufficient to impose that \(i\in I\) has no incentive to make a one-shot deviation from the equilibrium path. Following the same procedure as in the previous proofs one can easily show that \(\delta^\star_2\) exists in \((0,1)\) as long as the punishment stage payoff of investor \(i\in I\) is strictly lower than her on path stage payoff. The condition requires

\begin{equation*}
    -\frac{1}{2}(1-\alpha)^2(\hat{\sigma}^2_{1,\bar{P}}(1) + \sigma_2^2) + \frac{1}{2}\left[\frac{\alpha\beta}{1-\alpha}\right]^2>-\frac{1}{2}(1-\alpha)^2(\hat{\sigma}^2_{1,\bar{P}}(N) + \sigma_2^2) + \frac{1}{2}\left[\frac{\alpha\beta}{N-\alpha}\right]^2\frac{2N-1-\alpha}{1-\alpha},
\end{equation*}
or equivalently,
\begin{equation*}
    \hat{\sigma}^2_{1,\bar{P}}(1) -\hat{\sigma}^2_{1,\bar{P}}(N) < \left[\frac{\alpha\beta}{(1-\alpha)^2}\right]^2 - \left[\frac{\alpha\beta}{(N-\alpha)(1-\alpha)}\right]^2\frac{2N-1-\alpha}{1-\alpha}.
\end{equation*}
 which simply amounts to requiring that the policy influence gain thanks to the exercise of monopolistic market power is above the surprise loss due to less transparent communication than in the no-coordination case. Letting \(N\to\infty\) in the previous expression, it is verified that markets are better off in the monopolistic equilibrium relative to the efficient equilibrium if and only if \(\hat{\sigma}^2_{1,\bar{P}}(1) < \left[\frac{\alpha\beta}{(1-\alpha)^2}\right]^2\), which completes the proof.
\newpage

\section{Appendix: Silicon Valley Bank and Signature Bank Case Study}\label{appendix:svb}
In this Appendix, we provide a case study of Silicon Valley Bank (SVB) and Signature Bank (SB) to help illuminate a key idea in our theory: when the central bank is concerned about financial sector losses, it will underreact to inflationary pressure or output gap concerns (see Proposition \ref{underreac}). This underreaction will take the form of interest rates being lower than they would have been had the central bank not been concerned by such losses. In the language of our model, the interest rate will be lower than the economy-stabilizing rate (i.e., the rate needed for inflation/output gap stability).\footnote{Our theory also predicts that knowing the central bank will underreact, financial institutions will take greater risk. This case study is only intended to show an example of the central bank's underreaction rather than the consequent risk-taking decisions of financial institutions.}

SVB was a bank with over \$200 billion in assets and SB was a bank with over \$100 billion in assets. In March 2023, SVB faced severe problems as a result of interest rate risk, which it had actively not hedged (\citet{metrick2024}). The increases in interest rates combined with inadequate hedging negatively impacted its asset valuations which led to its failure when depositors began to withdraw their deposits. The failure of SVB led to widespread concerns and similarly rapid deposit outflows at SB which subsequently failed. These failures ultimately led to the `2023 Banking Crisis' and broader concerns around systemic risk.\footnote{\url{https://www.fdic.gov/news/press-releases/2023/pr23017.html}}

The pertinent question from the perspective of our theoretical model is whether these concerns around systemically risky banking sector losses induced the Fed to underreact to the high inflation during that period. Specifically, our theory predicts that the Fed set the fed funds rate (FFR) lower than it would have done in the counterfactual world of no banking crisis. However, this is difficult to prove given the empirical challenge of knowing that counterfactual.

We use three different methods to consider the counterfactual world: (1) Taylor rule predictions; (2) Fed Funds futures implied probabilities; and (3) FOMC minutes. Each of these support the idea that the Fed kept rates lower than it would have done in the absence of banking sector concerns and therefore are consistent with the predictions from our theoretical model. Below, we describe each method in more detail.
\\\\
\textbf{Taylor rule predictions}

The Taylor rule is an equation that prescribes a value for the FFR based on the values of inflation relative to target and the output gap (\citet{taylor1993}). In a sense, it captures the typical central bank loss function (the first term in equation \eqref{eq:cbloss}) and can be thought of as a theory-based prediction. The Taylor rule (and modifications of it) have been used consistently as a `benchmark' for the FFR. Therefore, one can compare the actual FFR to that predicted by the Taylor rule to see whether the Fed is keeping rates too low or too high.

We estimate the following Taylor rule\footnote{Note that we use a simple Taylor rule. One can use a number of different modified rules. For the purposes of our analysis, the different rules overall do not make a material difference.}:
\begin{align}
    \hat{r}_t
    &=
    \Bar{r}_t
    +
    \pi_t
    +
    \theta(\pi_t - \pi^*_t)
    +
    (1-\theta)\tilde{Y}_t\label{eq:taylor}
\end{align}
where $\hat{r}_t$ is the predicted FFR, $\Bar{r}_t$ is the long-run real interest rate, $\pi_t$ is inflation, $\pi_t^*$ is the target inflation rate, $\tilde{Y}_t$ is the output gap, and $\theta$ is the weight the central bank puts on inflation. We use commonly used values for the different variables. Specifically, we set the long-run real interest rate to 2\%, the target inflation rate to 2\%, and $\theta$ to 0.5.

Given the underlying data is quarterly, we can look at the predicted rate versus the actual rate in the second quarter of 2023 (the first data point after the Fed's rate-setting meeting). Our theory suggests that the Taylor rule predicted rate should be higher than the Fed's chosen rate because the former does not account for banking sector losses. Indeed, \eqref{eq:taylor} predicts a FFR of 6.17\% while the actual FFR was only 4.83\%. While this is consistent with our theory, the Taylor rule is typically not an accurate predictor of the FFR and therefore may not serve as effective counterfactual.
\\\\
\textbf{Fed Funds futures implied probabilities}

Next, we consider a market-based prediction of the FFR. Specifically, the probabilities of different rate decisions are computed using 30-day Fed Funds futures pricing data.\footnote{Our data is from the CME FedWatch Tool.} The useful feature of this method is that we can see the distribution of expectations rather than a single outcome.

Figure \ref{fig:FFR} below shows the probabilities of the different actions that the Fed will take according to market expectations. We look at the different probabilities from Friday 10 March to Friday 17 March. The important events are that between Friday 10 March and Sunday 12 March, both SVB and SB failed. Therefore, our theory would predict that FFR hikes should be less likely after that weekend than before it. This is precisely what we see. On Friday 10 March, the market priced in a 60\% chance of a 25 basis point hike, a 40\% chance of a 50 basis point hike, and a 0\% chance of no change. These were driven by the very high inflation at the time. However, just a few days later, there was a significant change in these probabilities. The probability of a 50 basis point hike fell to zero percent and stayed there while the probability of no change rose to over a third.
\begin{figure}[H]
    \centering
    \includegraphics[width=0.9\textwidth]{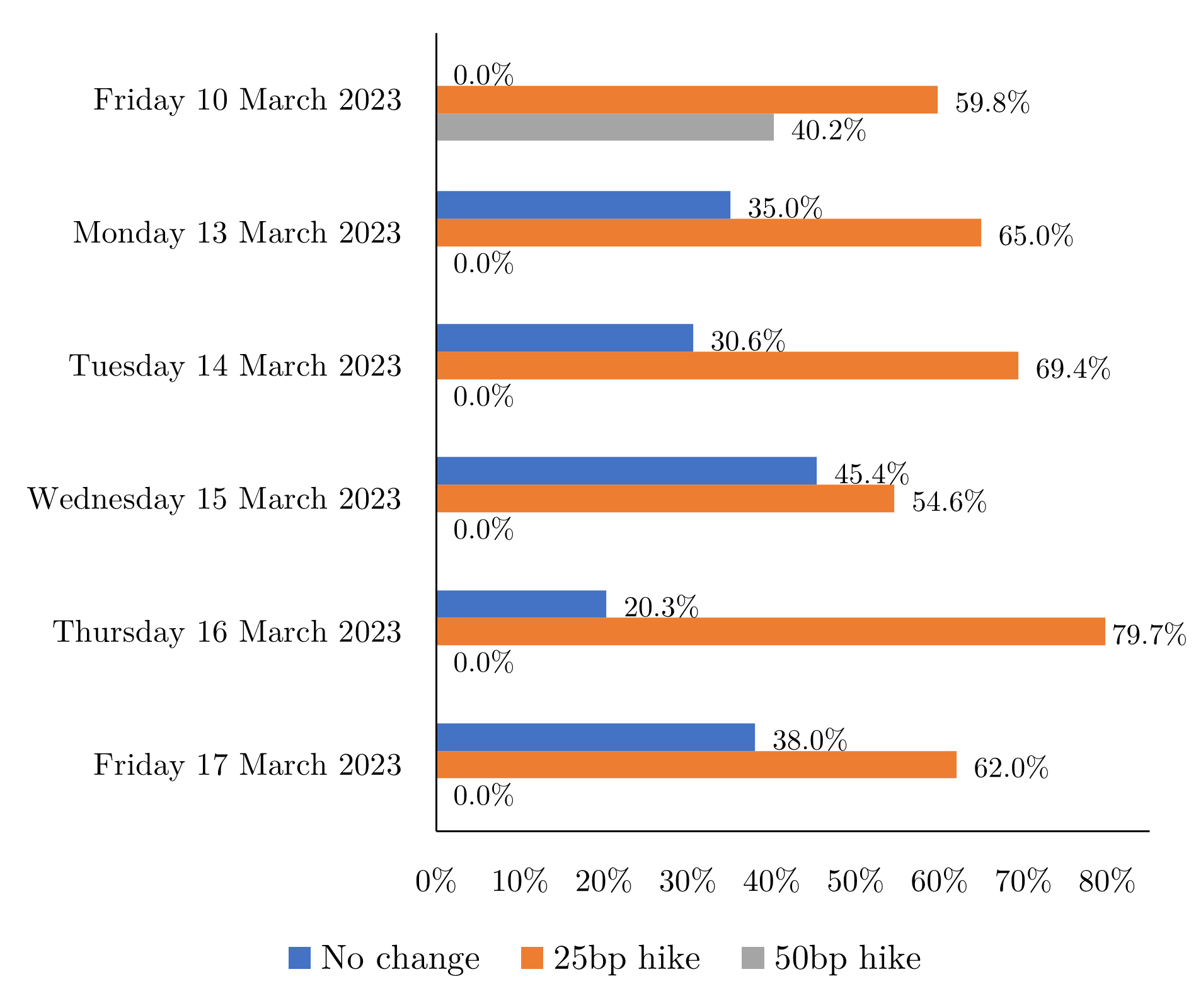}
    \caption{Probability of different FFR actions}
    \label{fig:FFR}
\end{figure}
The above Figure is consistent with our theory as inflation concerns did not suddenly disappear over that weekend. However, it is still possible, albeit unlikely, that the Fed may have been concerned about something else and that this timing is merely a coincidence.
\\\\
\textbf{FOMC Minutes}

Finally, to show that the Fed was explicitly concerned about banking sector developments and this played a role in its decision not to raise rates, we examine the minutes of the FOMC meeting in March 2023. The following excerpt from the minutes provide much clearer evidence of a counterfactual:

``Some participants noted that given persistently high inflation and the strength of recent economic data, they would have considered a 50 basis point increase in the target range to have been appropriate at this meeting \textit{in the absence of recent developments in the banking sector} [emphasis added].'' 

The Fed ultimately decided on a 25 basis point hike which is consistent with our theory that the Fed did not raise rates as much as it would have done had there been no banking sector concerns.

Ultimately, we believe that these three pieces of evidence show, through a leading example, how systemic stresses in the banking sector can lead that Fed to underreact to inflation or the output gap, consistent with our theoretical model.\footnote{Note that to have rigorous empirical evidence, one would need to conduct a systematic study of the data rather than a single case study.} 

\end{document}